\documentclass[universe,review,accept,pdftex,moreauthors]{Definitions/mdpi}

\renewcommand{\hl}{}
\firstpage{1}
\pubvolume{8}
\issuenum{6}
\articlenumber{323}
\pubyear{2022}
\copyrightyear{2022}
\externaleditor{Academic Editor: Alexander Andrianov} 
\datereceived{4 April 2022}
\dateaccepted{27 May 2022}
\datepublished{9 June 2022}
\hreflink{https://doi.org/10.3390/\linebreak universe8060323} 

\usepackage{url}
\def\UrlBreaks{\do\A\do\B\do\C\do\D\do\E\do\F\do\G\do\H\do\I\do\J
\do\K\do\L\do\M\do\N\do\O\do\P\do\Q\do\R\do\S\do\T\do\U\do\V
\do\W\do\X\do\Y\do\Z\do\[\do\\\do\]\do\^\do\_\do\`\do\a\do\b
\do\c\do\d\do\e\do\f\do\g\do\h\do\i\do\j\do\k\do\l\do\m\do\n
\do\o\do\p\do\q\do\r\do\s\do\t\do\u\do\v\do\w\do\x\do\y\do\z
\do\.\do\@\do\\\do\/\do\!\do\_\do\|\do\;\do\>\do\]\do\)\do\,
\do\?\do\'\do+\do\=\do\#}
\usepackage{tipa}
\Title{Lorentz Symmetry Violation of Cosmic Photons}

\TitleCitation{Lorentz Symmetry Violation of Cosmic Photons}

\Author{\hl{Ping He} $^{1}$\href{https://orcid.org/0000-0003-1935-4178}{\orcidicon} and Bo-Qiang Ma $^{1,2,3,}$*\href{https://orcid.org/0000-0002-5660-1070}{\orcidicon}} 

\AuthorNames{Ping He and Bo-Qiang Ma}

\AuthorCitation{He, P.; Ma, B.-Q.}

\address{%
$^{1}$ \quad School of Physics, Peking University, Beijing 100871, China; pinghe@stu.pku.edu.cn\\
$^{2}$ \quad Center for High Energy Physics, Peking University, Beijing 100871, China\\
$^{3}$ \quad Collaborative Innovation Center of Quantum Matter, \hl{Beijing 100871}, China}

\corres{Correspondence: mabq@pku.edu.cn}

\abstract{As a basic symmetry of space-time, Lorentz symmetry has played important roles in various fields of physics, and it is a glamorous question whether Lorentz symmetry breaks. Since Einstein proposed special relativity, Lorentz symmetry has withstood very strict tests, but there are still motivations for Lorentz symmetry violation (LV) research from both theoretical consideration and experimental feasibility, that attract physicists to work on LV theories, phenomena and experimental tests with enthusiasm. There are many theoretical models including LV effects, and different theoretical models predict different LV phenomena, from which we can verify or constrain LV effects. Here, we introduce three types of LV theories: quantum gravity theory, space-time structure theory and effective field theory with extra-terms. Limited by the energy of particles, the experimental tests of LV are very difficult; however, due to the high energy and long propagation distance, high-energy particles from astronomical sources can be used for LV phenomenological researches. Especially with cosmic photons, various astronomical observations provide rich data from which one can obtain various constraints for LV researches. Here, we review four common astronomical phenomena which are ideal for LV studies, together with current constraints on LV effects of photons.}


\begin{document}

\section{Introduction}\label{Introduction}

As a basic symmetry of space-time, Lorentz symmetry has played important roles in various fields of physics, and it can be ranked as the crystallization of human wisdom in understanding space-time. Since Einstein proposed special relativity in 1905, Lorentz symmetry has withstood the test of more than a hundred years in experiments with solid experimental foundation\footnote{But Ref. \cite{C1-cohen-2006-very} showed that many of the successful experiences of special relativity do not necessarily follow the theoretical framework of Lorentz symmetry.}. Then it is a natural question why we search for Lorentz symmetry violation (LV)? There are reasons from both theoretical consideration and experimental feasibility. From theoretical side, there is motivation for the research of new physics beyond relativity, such as quantum gravity, which may break Lorentz symmetry. From experimental side, even small violation of fundamental symmetry, such as LV, can lead to detectable effects at energy ranges well below the Planck scale $E_\text{Pl}$ with significant progresses in experimental techniques and methodologies for high precise measurements. Therefore, it is reasonable, necessary and feasible to study LV.

An early theoretical research on LV can be traced back to Dirac, who had attempted to reintroduce \ae ther into electrodynamics in 1951 \cite{C106-Dirac-1951-ather}. There are also other early perspectives related to LV \cite{C107-bjorken-1963-dynamical,C108-phillips-1966-graviton,C109-pavlopoulos-1967-breakdown}. From the origin, the theories including LV effects\footnote{It is worth noting that some theories are not born with the hope of destroying Lorentz symmetry, but the LV effects are inevitably produced in the development process, and we can judge these theories by these LV effects.} can be divided into three types: quantum gravity theory, space-time structure theory and effective theory with extra-terms.
\begin{itemize}
\item	Quantum gravity theory aims at unifying quantum theory and general relativity for building a unified basic theory. One typical representative quantum gravity theory is string theory (for a review see Ref. \cite{C21-gubser-2010-string}), which unifies quantum theory and general relativity by entering the concept of strings to solve the renormalization problem of gravitons. Besides string theory there are some other quantum gravity methods such as loop quantum gravity (for reviews see Refs. \cite{C124-rovelli-2008-loop,C168-Ashtekar-2021-short}), noncommutative quantum field theory (for a review see Ref. \cite{C125-szabo-2003-quantum}), analogue gravity (for a review see Ref. \cite{C112-barcelo-2011-analogue}), etc.
\item	When physicists consider whether space-time is not continuous but has the smallest basic unit, another type of LV theory was born. The thinking about the structure of space-time can be traced back to Pauli, who remarked that ``We may see herein an indication that not only the field concept, but also that the space-time concept in the microscale requires a principal modification\footnote{This sentence is originally in German: "Wir m{\"{o}}chten hierin einen Hinweis daf{\"{u}}r erblicken, dass nicht nur der Feldbegriff, sondern auch der Raum-Zeit Begriff im kleinen einer grunds{\"{a}}tzlichen Modifikation bedarf".}” after discussing the divergencies in quantum electrodynamics and the infinite self-energy of electrons in 1933 \cite{C140-Pauli-1933-handbuch}. Influenced by Pauli's idea, March pointed out the importance of a universal length in physics \cite{C119-march-1936-geometry,C120-march-1936-geometry,C121-march-1937-question,C122-march-1937-foundation}. Then Heisenberg \cite{C118-heisenberg-1938-universal}, de Broglie \cite{C141-debroglie-1943-die}, Pavlopoulos \cite{C109-pavlopoulos-1967-breakdown}, etc. also conducted theoretical explorations in this area. There are also theories starting from space-time geometry to seek possible space-time essences, e.g., Girelli, Liberati and Sindoni proposed a possible relation between modified dispersion relations and Finsler geometry to account for nontrivial structure of Planckian space-time \cite{C165-Girelli-2006-planck}. The discreteness of space-time also deserves scrutiny \cite{C158-Snyder-1946-quantized,C159-hohn-1957-on,C155-Xu-2011-universal,C156-Shao-2010-note,C157-Ma-2012-new}. The current theories about space-time structure and length units can be divided into two types: one introduces a length unit (maybe Planck length $L_\text{Pl}$) as a constant into the laws of physics, and a representative of that is doubly special relativity (DSR) \cite{C24-amelino-2001-relativity,C25-amelino-2010-DSR}. The other originates from the thinking about the nature of space-time itself: the space-time background metric can be affected by quantum fluctuations to behave like ``foam'' on short time and distance scale, and this model is called the space-time foam model, that is our current relatively unified understanding of the nature of space-time.
\item	In addition to the above two theories derived from basic principles, there are also extended models based on existing theories with extra-terms, and a representative is the extension of the standard model. Coleman and Glashow proposed the original version of an extension of the standard model by adding a LV term to the standard model Lagrangian \cite{C152-Coleman-1998-high}. Colladay and Kosteleck{\`y} developed a frame to deal with spontaneous CPT violation and Lorentz violation, then they obtained the CPT and LV terms \cite{C20-colladay-1997-CPT,C14-colladay-1998-lorentz}, and this extension model is called standard model extension (SME). Myers and Pospelov introduced higher dimensional LV operators which are not renormalizable \cite{C28-myers-2003-ultraviolet}. Such higher dimensional operators can naturally give the behaviors that are suppressed by high energy scales and seem to give a better mechanism than the renormalizable model, which needs to artificially make the LV coefficients tiny in order not to conflict with the existing experiments. Zhou and Ma proposed a framework with general requirement of physical independence or physical invariance of mathematical background manifold \cite{C43-zhou-2010-lorentz-1,C44-zhou-2011-theory-2,C153-zhou-2011-lorentz}. This framework introduces a background matrix $M^{\alpha\beta}$ by replacing the common derivative operators by the covariant derivative ones, leading to a Lorentz violation matrix for each type particle.
\end{itemize}

There are many theories including LV effects in the current frontier theoretical researches, and the LV phenomena predicted by different theories are not the same, so phenomenological and experimental studies of LV are urgently needed for identification. We should also mention that the classifications between quantum gravity theory and space-time structure theory are not so strict. In the development of many quantum gravity theories, physicists absorb the idea of space-time ``foam'', such as non-critical string theory~\cite{C90-ellis-1992-string,C91-ellis-1999-microscopic}, loop quantum gravity \cite{C92-gambini-1999-nonstandard,C93-alfaro-2002-loop}, string theory \cite{C94-kostelecky-1989-spontaneous}, doubly special relativity~\cite{C24-amelino-2001-relativity,C25-amelino-2010-DSR} and the effective field theory approaches \cite{C28-myers-2003-ultraviolet}. Some space-time structure theories are shown to be feasible of describing quantum gravity, e.g., Li and Chang constructed the theory of gravitation in Berwald–Finsler space \cite{C166-Li-2010-towards}, and Zhu and Ma got the arrival time delay of LV cosmological particles in the framework of Finsler geometry \cite{C167-zhu-2022-lorentz}.

We generally expect that LV effects have observable effects in extremely high-energy ranges. So before introducing these LV effects, we need to first introduce three energy scales, which are often used but often confused:
\begin{itemize}
\item	Planck scale $E_\text{Pl}$, at which Planck believed that when the strength of gravitational interaction and electromagnetic interaction are equivalent a new theory will emerge. Planck scale can be expressed by the energy dimension $E_\text{Pl}\equiv\sqrt{\hbar c^5 /G}\simeq 1.22\times 10^{28} \mathrm{eV}=1.22\times10^{19} \mathrm{GeV}$, the mass dimension $M_\text{Pl}\equiv E_\text{Pl}/c^2$ and the length dimension $L_\text{Pl}\equiv\sqrt{\hbar G/c^3}=\hbar c/E_\text{Pl}\simeq 1.6\times 10^{-35} \mathrm{m}$. Different dimensions are used depending on the theoretical expressing habits with same effect.
\item	Characteristic scale of the quantum gravity theory $E_\text{QG}$, at which the effects of a quantum gravity theory become significant.
\item	LV scale $E_\text{LV}$, at which the LV effects become significant. LV scale $E_\text{LV}$ can also mark the strength of the LV modification terms, and can be used in different order $E_\text{LV,n}$, where $E_\text{LV,1}$ (sometimes write as $E_\text{LV}$ for simplicity) is linear term and $E_\text{LV,2}$ is \mbox{quadratic term}.
\end{itemize}

We generally think that the LV modification terms can be suppressed by the integer power of the ratio of the typical energy $E$ of the actual physical process to the Planck scale $E_\text{Pl}$, so the linear LV energy scale $E_\text{LV}$ is near the Planck scale $E_\text{Pl}$. Generally, a quantum gravity theory will be realized near the Planck scale $E_\text{Pl}$, so the characteristic scale $E_\text{QG}$ is also near the Planck scale $E_\text{Pl}$. Although there is no one-to-one correspondence between a quantum gravity theory and a LV phenomenon, people often use the appearance of the LV phenomenon to mark a quantum gravity theory at work, so in a quantum gravity theory including LV effects, some people often use the characteristic scale $E_\text{QG}$ to replace the LV scale $E_\text{LV}$. To sum up, we expect that both the LV scale $E_\text{LV}$ and the quantum gravity characteristic scale $E_\text{QG}$ will appear near the Planck scale $E_\text{Pl}$, but their specific values need to be determined by experiments.

Currently, standard model and general relativity provide very successful descriptions of nature: general relativity describes gravity at the classical level, while standard model describes all other forces of nature down to the quantum level. However, there are irreconcilable contradictions between them, and there are some phenomena beyond the descriptions of the two. To reconcile the contradictions between standard model and general relativity, to explain some phenomena beyond existing theoretical frameworks, and to realize a complete and unified theoretical model, theoretical physicists have been searching for quantum gravity theories. These quantum gravity theories, which are expected to work at the Planck scale $E_\text{Pl}$, are very difficult to be verified in low-energy ranges. Fortunately, many of these quantum gravity theories predict various LV phenomena, that can be used to provide confirmation or falsification. On the other hand, the understanding and application of symmetry are great achievements of modern physics, and the discovery of symmetry violation is a treasure in the history of physics \cite{C5-lee-1956-question}. The discovery and experimental verification of important symmetry violation, such as parity violation and CP violation, have brought people a shocking awareness: is the Lorentz symmetry also an approximate symmetry in low-energy ranges? Considering the above two aspects, physicists are working with great enthusiasm on exploring the physical mechanisms, phenomenology and experimental explorations of LV.

In order not to contradict with existing experiments, any model that includes LV effects, regardless of the mechanisms, must ensure that the LV modifications are very small, and the observable phenomena can only be detected in the extremely high-energy ranges. However, the highest energy value of man-made particles on Earth today\hl{---}
$10^{13}\,\mathrm{eV}$ produced by the Large Hadron-Collider (LHC), and the highest-energy particles ever observed in cosmic-rays\footnote{Cosmic-rays are ultra-high-energy ($E\ge10^{20}\, \mathrm{eV}$) particles (mainly high-energy protons and other bare atomic nuclei such as helium nuclei and iron nuclei, also a small fraction of gamma-rays and neutrinos) emitted by distant active galaxies. The energy of cosmic-rays can reach $10^{21}\, \mathrm{eV}$ with unknown generation mechanism. The rays travel gigantic (cosmological) distances before reaching the Earth, then produce showers of elementary particles when pass through the Earth's atmosphere. There are mainly three kinds of observation methods: space observation, ground observation, and underground (or underwater) observation. It is widely believed that the study of cosmic-rays can yield a wealth of information about processes in most of the strange environments of the Universe, and the study of cosmic-rays has gradually become an important area of astrophysics researches.}\hl{---}$10^{21} \,\mathrm{eV}$ orders, are all many orders of magnitude lower than the Planck scale $E_\text{Pl}$. It can be foreseen that the LV effects are extremely small in terms of current experimental detection capabilities, and the LV effects are very difficult to be detected. Fortunately, the nature can “amplify” LV effects, such as the propagation of cosmological distances for cosmic-rays, etc. High-energy photons from the Universe are the highest-energy photons that people can receive\footnote{Beijing time on 17 May 2021, the Institute of High Energy Physics of the Chinese Academy of Sciences and Springer Nature jointly released the Large High Altitude Air Shower Observatory (LHAASO) major discovery: the highest energy of photon from the Universe---$1.42\, \mathrm{PeV}$ \cite{C88-cao-2021-ultrahigh}.}, and gigantic (cosmological) distance propagations can accumulate the LV effects. Current astronomical observation methods are also constantly improving, so a viable solution to seek signs of LV is from astronomical phenomenological researches, on which the current review is focusing.

By reviewing the current limitations of the photon LV phenomena and the theoretical models including LV effects, we state that different theoretical models predict different LV phenomena, and that the limitations from LV phenomena can verify or constrain different theoretical models. For clear illustration, Section \ref{Lorentz Violation Phenomena} briefly introduces some common LV phenomena. Section \ref{Theoretical Models including LV Effects} discusses three representative theoretical models and their predictions about LV phenomena, and Section \ref{Theoretical Models including LV Effects} introduces an effective model-independent method for LV researches. Section \ref{Four LV Phenomena of Photons} reviews and discusses four common photon LV phenomena and the current limitations from astronomical observations, then Section \ref{Four LV Phenomena of Photons} elaborates the validations and constraints of these limitations on the theoretical models.

\section{Lorentz Violation Phenomena}\label{Lorentz Violation Phenomena}

Our discussions of the experimental detections about LV effects are mainly focusing on the phenomena with the modified dispersion relation caused by LV and the resulting particle-specific behaviors\footnote{There are also some experiments to test the isotropic hypothesis of space, such as the clock experiment, etc.; this type of experiments belong to the laboratory experiments, but this paper mainly discusses the experimental tests of astronomical observations.}. Such LV phenomena mainly include (for reviews see \mbox{Refs. \mbox{\cite{C41-jacobson-2006-lorentz,C3-liberati-2009-lorentz}}):}
\begin{itemize}
\item	The light speed might depend on the energy or helicity of photons, then there will be vacuum dispersion and vacuum birefringence phenomena.
\item	Some inhibitory reactions in the standard model might occur, such as the decay of photons $\gamma\to e^\mathrm{+}+e^\mathrm{-}$, the splitting of photons $\gamma\to n\gamma$ and the electron pair emission $e^\mathrm{-}\to e^\mathrm{-}e^\mathrm{-}e^\mathrm{+}$.
\item \textls[-10]{The reaction characteristics in the standard model might be altered by the LV effects, thereby reflecting different behaviors from the standard model, such as the photon annihilation reaction $\gamma + \gamma_\mathrm{b} \to e^\mathrm{+}+e^\mathrm{-}$ might have some threshold anomalous behaviors.}
\item The different helicity particles might have different LV effects, such as an electron can reverse from one chiral state to another chiral state $e^\mathrm{\pm}\to e^\mathrm{\pm}+\gamma$.
\item The upper limitation on the velocity of particles, that is the maximum attainable velocity, will be changed by LV effects, such as the electron vacuum Cherenkov radiation $e^\mathrm{\pm}\to e^\mathrm{\pm}+\gamma$ might cause an upper limitation on the velocity of electron, etc.\\
\end{itemize}

Modification of the photon dispersion relation produces rich and measurable astrophysical phenomena. If LV causes the dispersion relation of photons to be no longer $E^2=p^2c^2$, but with a slight modification, then the speed of photon will no longer be a constant $c$. If the modification is related to the polarization state of photons, the photons with right-handed and left-handed polarization have different velocities, then we may observe vacuum birefringence. If the modification is related to the energy of photons, then high-energy photons may exhibit unusual physical phenomena: if the higher the energy, the slower the speed of light (subluminal), then we may observe the arrival time delay of high-energy photons and the threshold anomaly of photon annihilation reaction; if the higher the energy, the faster the speed of light (superluminal), then we may observe the decay of high-energy photons. Section \ref{Four LV Phenomena of Photons} discusses these four common astronomical phenomena.

Astrophysical measurements show outstanding advantages in the detection of LV phenomena. The highest-energy particles that people can currently receive come from the Universe, cosmological distance propagation can accumulate small LV effects to the measurable level, and current astronomical observation methods have partially achieved the accuracy of the requirements of the observations, so astrophysical measurements of cosmic-rays can provide sensitive detection of LV phenomena. Section \ref{Four LV Phenomena of Photons} discusses the limitations obtained from astrophysical measurements.

Various limitations obtained from experiments or observations can verify or constrain existing theoretical models that include the LV effects. Section \ref{Theoretical Models including LV Effects} discusses the verification and constraints on various theories by the currently obtained limitations. To better analyze different theories, we briefly introduce some common LV phenomena and the existing limitations before going into the detailed analyses.

\begin{itemize}
\item \paragraph{Arrival time delay of high-energy photons}

If the LV makes photons with higher energy slower (subluminal), then among photons departing from the same source at the same time, high-energy photons will arrive at the Earth slowly, and this phenomenon is called the arrival time delay of high-energy photons. Photon signals from the distant Universe can provide a reference time for measuring time delay phenomenon. The higher the photon energy and the longer the source distance, the more obvious the time delay effect and the tighter the limitation on this LV phenomenon. Amelino-Camelia et al. proposed that exploiting rapid changes in gamma-ray emission from distant astrophysical sources can be used to limit LV~\cite{C7-amelino-1998-tests}. Currently, LV has been tested by using observations of gamma-ray bursts (GRBs), active galactic nuclei (AGNs) and pulsars. There are studies suggesting positive signal for the light-speed variation while there are also studies proposed lower bound on the LV scale from data analyses.

\item \paragraph{Vacuum birefringence}

If the LV causes photons with different helicity to have different dispersion relations, that is, photons with right-handed and left-handed polarization have different velocities, then the polarization vector of the linear polarization plane will rotate. This rotation increases with the cosmological propagation distances of the photons and produces a measurable rotation angle, and this phenomenon is called vacuum birefringence. Since the polarization at the source is not known, the measurements of vacuum birefringence are inevitably affected by the inherent polarization angle. However, since the rotation angle caused by LV effects can offset partially, but not all, the polarization angle caused by emission mechanisms, the detection of polarized signals can still put an upper limitation on this possible LV effect. Currently, astronomical observations place strong limitations on vacuum birefringence, and these limitations lead to severe challenges to some of the theoretical models which allow vacuum birefringence.

\item \paragraph{Threshold anomaly of photon annihilation reaction}

If the LV modifies the dispersion relation of photons, the annihilation reaction between high-energy photons and low-energy photons can produce interesting physical phenomena. According to the special relativity, the photon annihilation reaction $\gamma + \gamma_\mathrm{b} \to e^\mathrm{+}+e^\mathrm{-}$ causes that photons with energy higher than the threshold are absorbed by low-energy photons, then the annihilation reaction prevents high-energy photons traveling long distances in the Universe, and the annihilation reaction results in the absorption modification of the spectrum, so we also call this reaction the background absorption of high-energy photons. However, if the LV produces a subluminal speed modification on the dispersion relation of photons, high-energy photons may not be absorbed by low-energy photons, so the high-energy photons can travel long distances and be received on Earth. The ultra-high-energy photons we have observed may be supporting evidences of this LV effect, and we will discuss that in detail later.

\item \paragraph{Decay of high-energy photons}

In the standard model, limited by the energy-momentum conservation, the photon decay reaction $\gamma\to e^\mathrm{+}+e^\mathrm{-}$ is prohibited, but in the theories including LV effects, photon decay may be a possible phenomenon \cite{C39-jacobson-2003-threshold}. If only the linear modification of the dispersion relation of photons is considered, when the modification is superluminal, the decay of photons will occur. If the LV modifies not only the dispersion relation of photons, but also the dispersion relation of electrons, then the decay of high-energy photons can also occur under the proper combination of modification parameters. At present, the observation data of LHAASO set very strict limitations on the decay of photons \cite{C81-14-li-2021-ultrahigh,C151-Li-2021-lhaaso,C66-lhaaso-2021-exploring, C82-chen-2021-strong}, and we will discuss that in detail later.

\end{itemize}

Besides photons, some other particles can also help LV researches, such as:

\begin{itemize}
\item \paragraph{Electron vacuum Cherenkov radiation}

Like photon decay, in some theories including LV effects, electrons may also decay. If the LV effects modify the dispersion relations of electrons and photons at the same time, the speed of electrons moving at extremely high speed may be greater than that of photons. Under the energy-momentum conservation, if the LV parameters of photons and electrons are suitable, there may be the decay of electrons $e^\mathrm{\pm}\to e^\mathrm{\pm}+\gamma$. It can also be analogous to charged particles propagating in a medium, when a charged particle moves through the medium, if its speed is greater than the light speed in the medium, it will radiate an electromagnetic field, and this phenomenon is called Cherenkov radiation. So the decay of electrons caused by LV can also be called electron vacuum Cherenkov radiation. If the LV effect causes the high-energy electron to decay, there will be no electron can reach the threshold of inverse-Compton reaction and accelerate photon to very high energy by inverse-Compton reaction in the Crab Nebula. However, recently LHAASO reported the detection of gamma-ray of energy up to $1.1 \mathrm{PeV}$ from the Crab Nebula \cite{C145-cao-2021-pata}, and Ref. \cite{C144-li-2022-testing} pointed out that this datum means severely strict limitation on the decay of electrons.

\item \paragraph{Chirality reversal of electrons}

If left-handed fermions and right-handed fermions have different LV effects (there is no theoretical reason for them to be equal), then the left-handed electrons and right-handed electrons can have different dispersion relations and LV parameters, so an electron can reverse from one chiral state to another chiral state, and this phenomenon is called chirality reversal of electrons $e^\mathrm{\pm}\to e^\mathrm{\pm}+\gamma$.

\item \paragraph{Upper limitation of electron synchrotron radiation}

The modified dispersion relations may cause anomalous behaviors on the upper limitation of the particle speed, especially in the process of synchrotron radiation of electrons, the LV effects are clearly displayed \cite{C42-jacobson-2003-strong,C160-Altschul-2006-limits}. In the theory with the Lorentz symmetry, the electron, performing synchrotron motion under the magnetic field $B$, has a maximum synchrotron frequency $\omega_\text{c}=3/2eB\gamma^3E$, where $\gamma$ is the Lorentz factor of the electron and $E$ is the energy of the electron. In the theory including LV effects, the maximum synchrotron frequency may be modified, so that the photons, emitted by synchrotron radiation, would also be affected. For the above considerations, there are some researchers studying the radiation of electrons in the Crab Nebula to get strict limitations on the electron LV parameters, such as Ref. \cite{C42-jacobson-2003-strong}.

\end{itemize}

Except photons and electrons, some other particles from the Universe can also be used to LV researches. Such as the earliest LV view was originated from “GZK paradox\footnote{Greisen, Zatsepin and Kuzmin predicted that limited by the reaction with the cosmic microwave background radiation, the cosmic-ray energies have a significant cutoff at $10^{20}\,\mathrm{eV}$, and this phenomenon is called Greisen–Zatsepin–Kuzmin (GZK) \mbox{cutoff~\cite{C115-greisen-1966-end,C116-zatsepin-1966-upper}}. Limited by the experimental conditions, the early experiments did not find an obvious cutoff of the cosmic-ray spectrum, and it led to the discussion of the “GZK paradox”. One explanation of the “GZK paradox” is that protons may have LV effects, and it is the origin of the LV phenomenon researches. With the improvement of experimental conditions, the existence of GZK-cutoff has been basically proved, and the ``GZK cutoff'' imposes strong constraints on the LV effects of protons, see, e.g., discussions in \mbox{Ref. \cite{C146-Xiao-2008-lorentz}}.}” of protons, and neutrino velocity modifications and oscillatory effects may also be related to the LV effects, etc. Since the photon observations are more mature, the most well-developed LV astronomical observational limitations are from photons.

It should be noted that not every LV theory has all of the above anomalous behaviors. Specific theories have their own specific LV behaviors, a specific theory reflects one or more of the characteristics or anomalies listed above, and we can verify or constrain these theories accordingly. Section \ref{Theoretical Models including LV Effects} briefly introduces the theoretical models including LV effects.

\section{Theoretical Models including LV Effects}\label{Theoretical Models including LV Effects}

As we have introduced, there are many kinds of theoretical models including LV effects: (i) Quantum gravity theory, which aims at solving the conflict between standard model and general relativity. One promising candidate quantum gravity theory is string theory; (ii) Space-time structure theory, which constructs new models from the perspective of space-time structure. One typical representative is doubly special relativity, which modifies the energy-momentum relation of special relativity; (iii) Effective theory with extra-terms. Since direct experiments near the Planck scale $E_\text{Pl}$ are impossible now, the outstanding problem on verifying Planck energy theories is the lack of experimental guidance. Because of this difficulty, effective field theory is a suitable tool for observing tiny signal of LV. One promising candidate is standard model extension \cite{C20-colladay-1997-CPT,C14-colladay-1998-lorentz}. In addition to these theories, we can also conduct LV studies by model-independent method.

Different theoretical models give different predictions of LV phenomena, generally, they give the modification of standard energy-momentum dispersion relation $E^2(p)=p^2+m^2$. However, the specific modification terms and modification behaviors are different depending on the specific theories, so the specific theories can be distinguished and identified by experiments and observations\footnote{Not every Plank energy theories have LV effects, nor does every theories including LV effects have modified energy-momentum dispersion relations.}.

This section discusses the three representative theories introduced above, and introduces a model-independent and effective method for LV researches. The discussion focuses on the different photon LV phenomena predicted by different theories and the current observation limitations on these theories.

\subsection{D-Foam Model in String Theory}\label{D-Foam Model in String Theory}

As the most promising approach to gravity quantization, string theory (for reviews see Refs. \cite{C21-gubser-2010-string,C170-Danielsson-2001-introduction}) promises to unify general relativity and quantum theory, promises to provide a unified theory that can describe all forces in nature, promises a new understanding of time, space, and additional dimensions yet to be discovered, and promises to provide a connection between some seemingly disparate concepts (such as black holes and quark-gluon plasma), so string theory is a very “promising” theory. String theory is at the forefront of all serious attempts to solve the mysteries hidden in fundamental physics, but there are many promises yet to be fulfilled.

String theory provides an elegant illation that begins with quantum theory and ends with general relativity. The difficulties of quantum gravity come from both technical problem and conceptual problem \cite{C170-Danielsson-2001-introduction}. The technical problem is related to non-renormalization, that is said the program of renormalization does not work in the quantum gravity case. String theory solves the non-renormalization by introducing the concept of strings, which are very small but have a definite length (a common view is $10^{-34}$ m). In string theory, everything is made out of string, including any measuring equipment, so it is impossible to measure anything smaller than a string. In string theory, it is meaningless to talk about distances smaller than the size of a string, so the non-renormalization problem is solved naturally. The conceptual problem is more complex, the quantization of space-time implies severe difficulties, such as an interesting problem is the black-hole information paradox, and string theory seems also to provide a solution to this conceptual problem.

String theory is a theory of continuous development. In the mid-1990s, the proposal of the brane concept prompted the second exploration of string theory \cite{C21-gubser-2010-string}. The proposal of the brane concept was affected by many aspects, such as: the interaction between strings becomes uncontrollable; and some symmetries, that appear in the supergravity theory\footnote{Supergravity theory is the low-energy limitation of superstring theory.}, are missing in the string theory, etc. A brane is like a string, but it can have any number of dimensions extending in space. A point particle is a 0-brane, a string is a 1-brane, a surface at any given time is a 2-brane, similarly 3-brane, 4-brane up to 9-brane. A D-brane is a special type of brane, that is where a string ends in space. In the development of string theory, people gradually realize that these higher-dimensional entities are not only examples of the richness of string theory, but they point towards deeper truths beyond string theory. String theory may not the last word of this theory, and it is very probable that strings play a quite subordinate role in the development of the theory \cite{C170-Danielsson-2001-introduction}. In the development of string theory, there are many new things, such as holography and M-theory, and we keep looking forward to the new development of string theory.

A strict challenge of string theory is the validation of this theory. Any real theory of nature must make predictions that can be verified through experiments or observations. String theory has two dramatic predictions: supersymmetry and extra dimensions \cite{C170-Danielsson-2001-introduction}. However, both predictions work at very high energy, and this property renders the string theory difficult to be verified through experiments or observations. Many string theorists believe that cosmology could be useful to the string theory validation, but it still needs time~\cite{C170-Danielsson-2001-introduction}. Since LV effects are predicted to work at Planck scale, the LV research may be useful to the validation of string theory. Our attention focuses on how to use LV research to verify the string theory.

In some certain string models, quantum gravitational fluctuations in the space-time background---“space-time foam”---can be treated as point-like D0-branes in the bulk space-time \cite{C83-ellis-2000-quantum,C84-ellis-2000-dynamical}, resulting that the vacuum might behave essentially like a dispersive medium that could violate Lorentz invariance. Refs. \cite{C86-13-li-2021-light,C150-Li-2021-light} worked in an anisotropic D-foam model~\cite{C87-ellis-2004-supersymmetric,C154-Ellis-2008-derivation,C155-Li-2009-time} and found that this supersymmetric string model for space-time foam has four characteristics: (i) This model results in a linear dependence between the speed of light and energy; (ii) This speed of light modification is subluminal; (iii) Photons are stable; (iv) Light velocity is helicity symmetric, so there is no vacuum birefringence. All these four characteristics correspond to current astronomical observations\footnote{Ref. \cite{C86-13-li-2021-light} pointed out that the light speed modification obtained from GRB data can be served as a key supporting evidence for this D-foam model, and since this model does not support the existence of photon decay and vacuum birefringence, it can naturally avoid very strong observational limitations from both phenomena.}. In this model, gauge bosons, such as photons, have modified dispersion relations. Current observational data from LHAASO can be used to explore the LV properties of photons and provide support for this model \cite{C81-14-li-2021-ultrahigh,C151-Li-2021-lhaaso}. In this model, charged particles, such as electrons, have dispersion relations consistent with the predictions of special relativity and are not subject to any (tree-level) quantum gravity modifications, and this property is supported by a new research~\cite{C144-li-2022-testing} which uses the latest observations of photons from the Crab Nebula by LHAASO. So this D-foam can avoid the strict constraints from the electron sector and thus be more viable.

Besides string theory, there are many approaches to quantum gravity, such as loop quantum gravity, whose key idea is the central lesson of general relativity: gravity is a manifestation of space-time geometry. Different ways to quantum gravity can cause different predictions of LV effects. D-foam model of string theory leads to a linear-order helicity independent effect in the modified dispersion relation of photon, and this property means that there is no vacuum birefringence to appear. In loop quantum gravity, the helicity dependence of modified dispersion is optional, there may be vacuum birefringence in a semiclassical regime that the electromagnetic field is a classical object whereas space is described by loop quantum gravity \cite{C92-gambini-1999-nonstandard}, and there may be no vacuum birefringence in other cases \cite{C93-alfaro-2002-loop}. Different predictions in different models can be tested by astronomical observational limitations.

\subsection{Doubly Special Relativity}\label{Doubly Special Relativity}

Amelino-Camelia gave the original discussions of doubly special relativity in \mbox{Ref. \cite{C24-amelino-2001-relativity}}, and he elaborated and explained some misunderstandings of this theory in \mbox{Ref. \cite{C25-amelino-2010-DSR}}. Some quantum gravity theories modify the energy-momentum dispersion relation near the Planck length $L_\text{Pl}$\footnote{$L_\text{Pl}$ is the Planck length, one kind of Planck scale. Planck length $L_\text{Pl}$ is a tiny length scale, and Planck length $L_\text{Pl}$ represent a new unified quantum gravity theory appears. Here, we follow the using method of Refs. \cite{C24-amelino-2001-relativity,C25-amelino-2010-DSR}.}, usually in the form of $E^2=p^2+m^2+\eta L_\text{Pl}^np^2E^2+\mathcal{O}(L_\text{Pl}^{n+1}E^{n+3})$. This modification leads to LV, and produces a class of preferred inertial observers (usually the same as the natural observers of the cosmic microwave background radiation). Amelino-Camelia proposed doubly special relativity to provide a scheme that does not require a preferred inertial reference frame \cite{C24-amelino-2001-relativity,C25-amelino-2010-DSR}.

The process of proposal of DSR is very similar to the emergence of special relativity. In Galileo relativity, there is no scale independent of the observer, and the energy-momentum relation is $E=p^2/(2m)$. While electromagnetic phenomena can be well represented by Maxwell equations, the fact, that these equations involve fundamental velocity scales\hl{---}light velocity $c$, seems to require a class of preferred inertial observers (that is said \ae ther). However, in the end, we found that there is no need for a preferred inertial reference frame but rather a new transformation law between inertial observers\hl{---}the Lorentz transformation. Special relativity introduces the first relativistic scale (light velocity $c$) independent of the observers, its dispersion relation takes the form $E^2=c^2p^2+c^4m^2$, and $c$ is now understood as a manifestation of the necessity of deforming the Galilean transformation in Maxwell equations. In DSR, one believes that the way out of quantum gravity may not require a new preferred inertial reference frame (maybe quantum gravity \ae ther), but need to revise the transformation laws between inertial reference frame, and the new transformation laws must contain two characteristic scales $c$ and $L_\text{Pl}$ rather than the single scale $c$ of ordinary special relativity. It is a reasonable conjecture: just as we set aside Galileo relativity (which does not have any characteristic invariant scale) in order to describe high-velocity particles and replace Galileo relativity with special relativity (which has an invariant velocity \mbox{scale $c$}); in order to describe ultra-high-energy particles, we may have to shelve special relativity and replace it with a new theory of relativity, maybe the DSR, which has two characteristic invariant scales: the original velocity scale $c$ and a new tiny length scale $L_\text{Pl}$ (same as a large energy scale $E_\text{Pl}$).

Currently, there are no unified mathematical formal descriptions of DSR, but some forms of ``DSR principles'' can still be summarized \cite{C24-amelino-2001-relativity,C25-amelino-2010-DSR}:
\begin{itemize}
\item	(RP): The physical laws are the same in all inertial frames (for all inertial observers). Particularly, the parameters that appear in the physical laws take the same values in all inertial frames. Equivalently, if two inertial observers in relative motion setup the same experimental procedures they get exactly the same numerical values (same dimensions) for the measurement results. This principle actually states that Galileo relativity is valid.
\item	(La): The physical laws, in particular the transformation laws between inertial observers, involve a fundamental/observer-independent small length scale $L_\text{dsr}$ (possibly Planck length $L_\text{Pl}$), which can be measured by each inertial observer following the measurement procedure $\mathcal{M}_{L_\text{dsr}}$. This principle actually requires that DSR must have a principle that giving the length scale $L_\text{dsr}$, or the corresponding momentum/energy/frequency scale $1/L_\text{dsr}$.
\item	(Lb): The physical laws, in particular the transformation laws between inertial observers, involve a fundamental/observer-independent velocity scale $c$, which can be measured by each inertial observer as the light velocity with wavelength $\lambda$ much larger than $L_\text{dsr}$, more rigorously, $c$ is obtained as the infrared $\lambda/L_\text{dsr}\to \infty$ limit of the speed of light.
\end{itemize}

The principle (La) is clearly incomplete: we do not have enough experimental information and even cannot make a description of the $\mathcal{M}_{L_\text{dsr}}$ procedure, which measure $L_\text{dsr}$. Amelino-Camelia also gave an alternative to the principle (La) related to the dispersion relations \cite{C25-amelino-2010-DSR}:
\begin{itemize}
\item	(La*): The physical laws, in particular the transformation laws between inertial observers, involve a fundamental/observer-independent small length scale $L_\text{dsr}$ (possibly Planck length $L_\text{Pl}$), which can be measured by each inertial observer by determining the dispersion relation for photons. \textls[-15]{This relation has the form \mbox{$E^2-c^2p^2+f(E,p;L_\text{dsr})=0$}}, where the function $f$ is same for all inertial observers, particularly all inertial observers agree on the leading $L_\text{dsr}$ dependence of $f: f(E,p;L_\text{dsr})\simeq L_\text{dsr}cp^2E$.
\end{itemize}

DSR can be verified or constrained by phenomenological research \cite{C24-amelino-2001-relativity,C25-amelino-2010-DSR}, such as: the decay of photons, dependence of speed of light on wavelength (energy), threshold anomalies for photon annihilation reactions, the research of the Crab Nebula synchronous cyclotron radiation data, etc. At present, there are many explorations on the mathematical form of DSR, and different mathematical realization forms have different tolerances for LV phenomena. For example, a DSR model latest proposed by Li and Ma can cover the subluminal energy dependence of light speed \cite{C123-li-2022-model,C142-li-2022-doubly}, and this model can withstand the strict limits both on vacuum birefringence and photon decay. This model does not produce threshold anomalies for photon annihilation reactions, while some other models \cite{C117-amelino-2003-kinematical} allow threshold anomalies in photon annihilation processes. It suggests us that even with the same theoretical thought, there may be differences in the LV effects and the LV phenomena brought about by different mathematical realizations. It requires us to discriminate and deal with each model carefully.

DSR is actually a space-time theory, and the LV effects that DSR contains are actually generated by the replacement of Lorentz symmetry with higher-level symmetry. For example, in DSR, the energy-momentum conservation does not disappear but is just no longer the simple form $E^2=p^2+m^2$ in the ordinary sense, and the ordinary energy-momentum conservation is replaced by unknown higher-level form. In addition to phenomenological researches, other aspects of DSR research are also being carried out gradually, and whether this theory is correct or not and its future development directions still need more in-depth researches.

\subsection{Standard Model Extension}\label{Standard Model Extension}

The standard model extension is to introduce a tiny LV modification on the basis of the standard model, that is, the total Lagrangian is:
\begin{linenomath}
\begin{equation}\label{SME-L}
\mathcal{L}=\mathcal{L}_\mathrm{SM}+\delta\mathcal{L}.
\end{equation}
\end{linenomath}

SME contains standard model and all operators that may break Lorentz symmetry. SME can be divided into two parts: (i) The minimal extension, whose operator dimension $d\le4$; (ii) The non-minimal extension, whose operators have higher dimension. Ref. \cite{C20-colladay-1997-CPT} and Ref. \cite{C14-colladay-1998-lorentz} systematically gave the CPT violation minimal extension of standard model, and this extension is renormalizable. In spite of the non-renormalizability, the advantage of introducing LV modification terms in higher mass dimensions is obvious, as these higher dimensional operators can give the high energy scale (maybe Planck scale $E_\text{Pl}$) depression behaviors naturally, and they give the mechanism of the depression of the Planck scale better.

Besides SME, there are also many other ways to extend the standard model. Before we go into the details of SME, we introduce a simple extension model---the Coleman--Glashow model \cite{C152-Coleman-1998-high}.

\subsubsection{The Coleman--Glashow Model}\label{The Coleman--Glashow Model}

The Coleman--Glashow model \cite{C152-Coleman-1998-high} is ranked as the most simple and original version of an extension of the standard model in the effective field theory framework \cite{C143-Ma-2012-new}. In the Coleman--Glashow model, the Lagrangian with LV term is:
\begin{linenomath}
\begin{equation}\label{SME-CG}
\mathcal{L}=\mathcal{L}_\mathrm{SM}+\partial_i\Psi\epsilon\partial^i\Psi,
\end{equation}
\end{linenomath}
where $i=1,2,3$, $\Psi$ denotes a set of $n$ complex scalar fields assembled into a column vector, and $\epsilon$ is a Hermitian matrix that signals the LV effect in the Coleman--Glashow model. For fermions, the wave function $\Psi$ denotes the Dirac spinor, and the LV parameter $\epsilon$ can be taken as a fixed scalar constant for the fermion under consideration.

The Coleman--Glashow model cannot be considered as an “exact” theory \cite{C143-Ma-2012-new}. The reason is that the Lagrangian Equation (\ref{SME-CG}) cannot be taken as invariant in all inertial frames of reference, but only valid in one inertial reference frame where the observer is working. When the observer is focusing on the LV effect within a certain frame but does not care about relations between different frames, the observer can adopt the Coleman--Glashow model. Otherwise, with the requirement of physical event consistency \cite{C143-Ma-2012-new}, the situation could become very complicated: there will be different formats in different frames.

Physically, the LV effect of the Coleman--Glashow model \hl{comes} from the existence of a “background” $\epsilon$ \cite{C143-Ma-2012-new}. In the Coleman--Glashow model, the parameter $\epsilon$ for a certain particle keeps as a constant without change in difference reference frames. This property is different from the SME, and we will explain the LV source of SME later. 

The Coleman--Glashow model can apply as a practical tool to demonstrate or estimate the magnitude of the LV effect. It has been successfully applied in many phenomenological analyses to constrain the LV effect in some physical processes. For example, \mbox{Ref. \cite{C146-Xiao-2008-lorentz}} adopted this model to analyze the constraint on LV effect of the proton from the observation of the GZK cutoff and Ref. \cite{C149-Bi-2011-constrains} used this model in the research to constrain neutrino LV~effect.

\subsubsection{Minimal Standard Model Extension}\label{Minimal Standard Model Extension}

Colladay and Kosteleck{\`y} developed a frame to deal with spontaneous CPT violation and Lorentz violation\footnote{CPT violation definitely leads to Lorentz violation, but Lorentz violation does not mean CPT violation.} in the context of conventional effective field theory \cite{C20-colladay-1997-CPT,C14-colladay-1998-lorentz}, and the minimum CPT and Lorentz violation terms are obtained in this frame. Their approach is very intuitive and effective: they took the expected vacuum value (which is due to the spontaneous Lorentz violation) of the tensor field, as the constant coefficient in the Lagrangian, and they introduced the coupling between the expected vacuum value and the field quantities of the standard model, then the coupling shows the LV effects. In the pure photon sector we are concerned with, the Lagrangian is \cite{C14-colladay-1998-lorentz}:
\begin{linenomath}
\begin{equation}\label{Min SME-L}
\mathcal{L}_\mathrm{SME}=-\frac{1}{4}F^{\mu\nu}F_{\mu\nu}+\mathcal{L}_\mathrm{photon}^\mathrm{CPT-even}+\mathcal{L}_\mathrm{photon}^\mathrm{CPT-odd},
\end{equation}
\end{linenomath}
where $\mathcal{L}_\text{photon}^\text{CPT-even}$ is even under the CPT transform, reading:
\begin{linenomath}
\begin{equation}\label{Min SME-L(even)}
\mathcal{L}_\mathrm{photon}^\mathrm{CPT-even}=
-\frac{1}{4}(k_\mathrm{F})_{\kappa\lambda\mu\nu}F^{\kappa\lambda}F^{\mu\nu},
\end{equation}
\end{linenomath}
and $\mathcal{L}_\text{photon}^\text{CPT\text{-}odd}$ is odd under the CPT transform, reading:
\begin{linenomath}
\begin{equation}\label{Min SME-L(odd)}
\mathcal{L}_\mathrm{photon}^\mathrm{CPT-odd}=\frac{1}{2}(k_\mathrm{AF})^{\kappa}\epsilon_{\kappa\lambda\mu\nu}A^{\lambda}F^{\mu\nu}.
\end{equation}
\end{linenomath}
Coefficients $(k_\text{F})_{\kappa\lambda\mu\nu}$ and $(k_\text{AF})^{\kappa}$ are CPT-even and CPT-odd respectively.

From the Lagrangian, SME actually introduces some “background” fields, which are transformed between different reference frames in the form of tensors. The Lorentz violation is due to the existence of the “background” fields. The standard model particles exhibit LV effect at a certain frame of the observer by taking these ``background'' fields as fixed. From a strict sense, there is no Lorentz violation for the whole system of the standard model particles together with the “background” fields \cite{C143-Ma-2012-new}.

The minimal SME preserves various desirable features of standard quantum field theories such as gauge invariance, energy-momentum conservation, observer Lorentz invariance, Hermitian property, validity of conventional quantization methods, and power-counting renormalizability \cite{C14-colladay-1998-lorentz}. Other important features such as positivity of energy, micro-causality, and the general anomaly eliminations are also expected \cite{C14-colladay-1998-lorentz}. However, both dimensionality problem and the implications of spontaneous LV of gravity at observable energy ranges have not been resolved \cite{C14-colladay-1998-lorentz}.

\textls[-20]{The minimal SME can be verified or constrained through the observations of some experimental phenomena. The CPT violation terms allowed by SME have two properties\hl{---}odd and even}, the minimal SME predicts that the light velocity is modified to energy at the first-order, and the minimal SME allows superluminal and subluminal phenomena at the same time. So, in photon sector, minimal SME produces the delay of high-energy photons, vacuum birefringence, the threshold anomaly of photon annihilation reaction and the decay of high-energy photons, etc.

\subsubsection{Non-Minimal Standard Model Extension}\label{Non-minimal Standard Model Extension}

Myers and Pospelov pointed out that under the framework of effective field theory, it is also possible to introduce higher dimensional LV operators, which are not renormalizable~\cite{C28-myers-2003-ultraviolet}. The benefit of introducing LV modification terms in higher mass dimensions is that the higher dimensional operators can naturally give the behaviors that are suppressed by high energy scales (maybe Planck scale $E_\text{Pl}$), such as coefficients proportional to $1/E_\text{Pl}$, $1/E_\text{Pl}^2$, etc. In this way, this non-renormalizable model seems to give a better mechanism than the renormalizable model, which needs to artificially make the LV coefficients tiny in order not to conflict with the existing experiments.

In this model, due to the existence of the vacuum expectation value of the tensor field, it is inevitable to introduce a special preferred frame, which is marked by the background 4-vector $n^a$ ($n\cdot n=1$). Consider the gauge field (i.e. photon field) that we care about, through the framework of Myers and Pospelov \cite{C28-myers-2003-ultraviolet}, with the original standard model Lagrangian $\mathcal{L}_\text{0}=-(F_{\mu\nu}F^{\mu\nu})/4$, the 5-dimensional non-renormalizable LV modification is~\cite{C28-myers-2003-ultraviolet}:
\begin{linenomath}
\begin{equation}\label{non Min SME-L}
\mathcal{L}_\gamma=\frac{\xi}{M_\mathrm{Pl}}n^aF_{ad}n\cdot\partial(n_b\widetilde{F}^{bd}),
\end{equation}
\end{linenomath}
where $\widetilde{F}^{ab}=1/2\epsilon^{abcd}F_{cd}$, and this modification is odd under the CPT conjugation when it is even under the charge conjugation. After the introduction of the LV modifications, the equation of motion becomes:
\begin{linenomath}
\begin{equation}\label{non Min SME-motion equation}
\Box A_a=-\frac{\xi}{M_\mathrm{Pl}}\epsilon_{abcd}n^b(n\cdot\partial)^2F^{cd}.
\end{equation}
\end{linenomath}

For the photon with 4-momentum $k^a=(E,0,0,p)$, the energy-momentum dispersion relation becomes:
\begin{linenomath}
\begin{equation}\label{non Min SME-dispersion relation}
(E^2-p^2\pm\frac{2\xi}{M_\mathrm{Pl}}p^3)(\epsilon_\mathrm{x}\pm i\epsilon_\mathrm{y})=0.
\end{equation}
\end{linenomath}

With the improvement of experimental precision, it was found that the linear modifications are not enough to explain the Lorentz invariance observed in experiment. In this case, one can look for higher-dimensional modification terms. Mattingly gave the LV modifications of the 6-dimensional mass dimension in the case of spatial rotation invariance \cite{C2-mattingly-2008-have}. Compared to 5-dimensional operators, higher dimensional operators can better explain the smaller depression of the Planck scale \cite{C2-mattingly-2008-have}. For the considerations of renormalize group evolution and the conjecture of CPT conservation, the 6-dimensional LV modifications seem to be more natural\footnote{The introduction of supersymmetry is needed to avoid the higher-dimensional LV operators producing the behavior of low-dimensional LV operators \cite{C35-bolokhov-2005-lorentz,C36-nibbelink-2005-lorentz}.}. There are many more studies and discussions on 5-, 6- and higher dimensional operators, such as Refs. \cite{C63-xiao-2009-constraints,C146-Xiao-2008-lorentz} obtained modified dispersion relations of minimal SME and 5-d SME by requiring the vanishing of determinants of inverse of modified photon propagators.

Besides minimal SME and non-minimal SME, there are also other ways of extending the standard model.
For example, Zhou and Ma proposed a framework, with general requirement of physical independence or physical invariance of mathematical background manifold \cite{C43-zhou-2010-lorentz-1,C44-zhou-2011-theory-2,C153-zhou-2011-lorentz}. This framework introduces a background matrix $M^{\alpha\beta}$ by replacing the common derivative operators by the covariant derivative ones, and can be referred as standard model supplement or Lorentz violation matrix theory, as the Lorentz violation effect is contained in a Lorentz violation matrix for each type particle. There are also many people starting from SME, extending other theoretical and phenomenological studies in different fields, and providing constraints for SME, so the SME is under constant research and refinements.

In SME, the LV parameters are particle-dependent and process-independent. This property means that the LV parameters of different particles are different, but for a specific parameter of a specific particle, the value is not affected by the specific process. For example, for the modification parameter of dispersion relation, the parameter of photons and electrons are different. However, only for photons, the parameters that predict the time delay effect also predict the vacuum birefringence effect. This property means that SME predicts time delay effect and vacuum birefringence effect at the same time, and the strict limitations on vacuum birefringence effect also impose strict restrictions on the time delay effect. However, this process-independent property does not hold for all theoretical models, as we have already shown, D-foam model in string theory predicts the time delay effect but no vacuum birefringence effect.

From the above discussion we can clearly see that different theoretical models have different LV effects and predict different LV phenomena. This property requires us to be very careful when dealing with LV parameters which are obtained from different processes~\cite{C81-14-li-2021-ultrahigh}. As different theoretical models have different LV effects and predict different LV phenomena, we can use the limitations obtained from specific LV phenomenon to verity or constrain different theoretical models. Such as \mbox{Ref. \cite{C28-myers-2003-ultraviolet}} pointed out that the strict limitations derived from the vacuum birefringence might set strict limitations on the framework of the SME with 5-d CPT-odd operators.

In the next section, we detail the currently obtained astronomical observational limitations for four common photon LV phenomena, as well as their verification or constraints on theoretical models. Before that we need to introduce a model-independent method.

\subsection{Model-Independent Method}\label{Model-Independent Method}

Since different theories are based on different mechanisms, there are even conflicts between them. To describe and study the LV physics more conveniently, we adopt a model-independent method. Regardless of the specific theories, we assume that the general form of the dispersion relation is: $E^2=f(p;E_\text{LV},\{\xi_\text{i}\})$, where $f(p;E_\text{LV},\{\xi_\text{i}\})$ is a general function of photon momentum $p(p=|\vec{p}|)$ and a set of parameters $\{\xi_\text{i}\}$ \cite{C63-xiao-2009-constraints}, then we perform Taylor expansion to get:
\begin{linenomath}
\begin{equation}\label{model indep-Taylor expansion}
E^2\simeq p^2c^2[1-\sum_{n=1}^{\infty}s_\mathrm{\pm}(\frac{|\vec{p}|}{E_\mathrm{LV,n}})^n],
\end{equation}
\end{linenomath}
where $E_\text{LV,n}$ is the hypothetical energy scale at which the $n$th-order LV effect becomes significant, and $n=1$ and $n=2$ are the linear and quadratic energy dependence. $s_\mathrm{\pm}$ is the theoretically relevant factor, $s=+1$ means that the higher the energy the slower the speed of light (subluminal), and $s=-1$ means faster (superluminal).

Model-independent method can adopt different parameterization according to different research needs. When the characteristic energy of the research process is much smaller than the LV scale $E_\text{LV}$, the summation is dominated by the leading-order. We assume that LV brings about the leading-order $(|\vec{p}|/E_\text{LV,n})^n$ but no influence of other orders, then \mbox{we get}:
\begin{linenomath}
\begin{equation}\label{model indep-leading order}
E^2=p^2c^2[1-s_\mathrm{n}(\frac{|\vec{p}|}{E_\mathrm{LV,n}})^n],
\end{equation}
\end{linenomath}
and the light velocity modification is:
\begin{linenomath}
\begin{equation}\label{model indep-leading order(light velocity)}
v=\frac{\partial E}{\partial p}\approx c[1-s_\mathrm{n}\frac{n+1}{2}(\frac{|\vec{p}|}{E_\mathrm{LV,n}})^n].
\end{equation}
\end{linenomath}

In this parameterization method, one can only consider the leading-order influence with the most prominent contribution but ignore the influence of other orders, so this parameterization method can make the phenomenological research more convenient.

When we consider both linear and quadratic modifications, we can use the following parameterization:
\begin{linenomath}
\begin{equation}\label{model indep-linear and quadratic(light velocity)}
v(E)=c(1-\xi\frac{E}{M_\text{Pl}c^2}-\zeta\frac{E^2}{M_\text{Pl}^2c^4}),
\end{equation}
\end{linenomath}
where $M_\text{Pl}$ is the Planck mass scale, and $\xi,\zeta$ are dimensionless parameters. This parameterization method conventions are also commonly used.

Model-independent method facilitates phenomenological researches. Using model-independent method allows us to temporarily ignore the conflicts between different theoretical models. Model-independent method can be applied to the study of the arrival time delay of high-energy photons, vacuum birefringence, the threshold anomaly of photon annihilation reaction, and the decay of high-energy photons, etc.

In model-independent method, the LV parameters are particle-dependent and process-dependent, that is to say, the parameters of different processes of different particles may be different. Since different theoretical models have different LV effects and predict different LV phenomena, we need to be very careful when dealing with LV parameters which are obtained from different processes \cite{C81-14-li-2021-ultrahigh}, and the model-independent method provides the convenience of such researches. Such as, we can set different parameters when we deal with time delay effect and vacuum birefringence effect, and the values of the specific LV parameters need to be judged by experiments. For the physical meanings behind the parameters, we need further researches.

Section \ref{Four LV Phenomena of Photons} will adopt model-independent method in the exposition of LV phenomenon researches. Using this method means that we approach LV physics from phenomenology, and we assume that the parameterization of different particles is particle-dependent and process-dependent. We do not discuss the physical meanings behind the LV parameters, since deeper physical meanings require more in-depth theoretical researches.

\section{Four LV Phenomena of Photons}\label{Four LV Phenomena of Photons}

To verify or constrain the LV theories, what we need now are the supports from experimental observations. Although the observable LV phenomena are tiny at observable energy ranges, the tiny LV phenomena can accumulate to measurable levels over gigantic (cosmological) distances; therefore, sensitive detections of LV can be provided by astrophysical measurements involving long distance propagations. This section will introduce the mechanisms and the research status of the four photon LV phenomena: the arrival time delay of high-energy photons, vacuum birefringence, the threshold anomaly of photon annihilation reaction and the decay of high-energy photons.

\subsection{Arrival Time Delay of High-Energy Photons}\label{Arrival Time Delay of High-Energy Photons}

Vacuum dispersion is a potential LV phenomenon. If the LV effects make photons with higher energy traveling slower (subluminal), then among photons departing from the same source at the same time, high-energy photons will arrive at the Earth more slowly, and this phenomenon is called the arrival time delay of high-energy photons. Regardless of the specific theories, we can use the LV photon modified dispersion relation based on the model-independent method:
\begin{linenomath}
\begin{equation}\label{time delay-dispersion relation}
E^2\simeq p^2c^2[1-\sum_{n=1}^{\infty}s_\mathrm{\pm}(\frac{E}{E_\mathrm{LV,n}})^n],
\end{equation}
\end{linenomath}
where $E_\text{LV,n}$ is the hypothetical energy scale at which the $n$th-order LV effect becomes significant. $s_\mathrm{\pm}=\pm$ is the theoretically relevant factor. When the physical process energy $E$ is much smaller than the LV scale, i.e., $E\ll E_\text{LV}$, the summation is dominated by the leading-order in the sequence. Considering only the leading-order, we can get the photon velocity:
\begin{linenomath}
\begin{equation}\label{time delay-light velocity}
v(E)=\frac{\partial E}{\partial p}\approx c[1-s_\mathrm{\pm}\frac{n+1}{2}(\frac{E}{E_\mathrm{LV,n}})^n],
\end{equation}
\end{linenomath}
where $n=1$ and $n=2$ are the linear and quadratic energy dependence. $s=+1$ means that the higher the energy the slower the speed of light (subluminal), and $s=-1$ means faster the light velocity (superluminal). Because of the energy dependence of $v(E)$, two photons, which simultaneously emit from the same astrophysical source at redshift $z$ and with different observer coordinate system energies $E_\text{h}>E_\text{l}$, will arrive at the observer at different times. Considering the expansion of the Universe, we can get the expression for the LV-induced time delay \cite{C128-jacob-2008-lorentz}:
\begin{linenomath}
\begin{equation}\label{time delay-LV delay}
\Delta t_\mathrm{LV} =t_\mathrm{h}-t_\mathrm{l}=s_\mathrm{\pm}\frac{1+n}{2H_\mathrm{0}}\frac{E_\mathrm{h}^n-E_\mathrm{l}^n}{E_\mathrm{LV,n}^n}\int_0^z\frac{(1+z')^ndz'}{\sqrt{\Omega_\mathrm{m}(1+z')^3+\Omega_{\Lambda}}},
\end{equation}
\end{linenomath}
where $t_\text{h}$ and $t_\text{l}$ represent the arrival times of high-energy and low-energy photons respectively. $H_\text{0}$ is the Hubble constant, and $\Omega_\text{m}$ and $\Omega_{\Lambda}$ are the matter energy density and vacuum energy density (here we use the flat $\Lambda\mathrm{CDM}$ Universe model). Since $E_\mathrm{h}\gg E_\mathrm{l}$, it can be approximated as $E_\mathrm{h}^n-E_\mathrm{l}^n\approx E_\mathrm{h}^n$.

If we adopt $z=20$ as the fixed upper limitation for the redshift of any source, the linear LV delay has the limitation: $|\Delta t_\mathrm{LV} |\le0.5(E_\mathrm{h,GeV}/\zeta)s$ and the quadratic case is $|\Delta t_\mathrm{LV}|\le5.2\times10^{-19} (E_\mathrm{h,GeV}/\zeta)^2s$, where $E_\mathrm{h,GeV}=E_\mathrm{h}/(1 \mathrm{GeV})$, $\zeta=E_\mathrm{LV}/E_\mathrm{Pl}$ \cite{C50-wei-2021-testing}. It means that the linear terms can cause potentially observable delay, while the quadratic terms are so tiny that they are impossible to be observed with the time-of-flight technique \cite{C50-wei-2021-testing}.

Photon signals from the distant Universe can provide a reference time for measuring time delay effect. Equation (\ref{time delay-LV delay}) shows that the time delay caused by LV increases with high-energy photon energy and source distance, so the higher the photon energy and the longer the source distance, the tighter the limitations on the LV scale. Amelino-Camelia et al. proposed exploiting rapid changes in gamma-ray emission from distant astrophysical sources to limit LV \cite{C7-amelino-1998-tests}. Currently, time delay effect has been tested using observations of gamma-ray bursts, active galactic nuclei and pulsars. Finally, we need to mention that the time delay effect has other explanation: the dispersive properties of the interstellar medium, this explanation maintains Lorentz symmetry, and this explanation needs more research \cite{C161-tonks-1929-oscillations,C162-Jackson-1999-classical,C163-Dobrynina-2014-photon,C164-Brevik-2020-dispersion}.

\subsubsection{Gamma-ray Bursts}\label{Gamma-Ray Bursts}

Gamma-ray bursts (for a review see Ref. \cite{C49-piran-2005-GRB}) are the most energetic astrophysical processes except the Big Bang. The energies of cosmic-rays span a wide range: (i) Ultra-high-energy cosmic rays: their energy range from $10^{20} \mathrm{eV}$ or higher; (ii) The energies of cosmic photons from GRBs can reach $10\sim100 \mathrm{GeV}$ or higher; (iii) Cosmic neutrinos: they have higher energies at the order of $\mathrm{TeV}\sim\mathrm{PeV}$. GRBs are generally divided into two types---long bursts and short bursts. Long GRBs with a duration greater than $2$ s may come from the collapse of massive rapidly rotating stars. Short GRBs with a duration less than \mbox{$2$ s} may come from the coalescence of two neutron stars or a neutron star with a black hole. The sources of GRBs are far away from the Earth. Generally, the mean redshift of a long GRB is about $z\approx2.15$, which is equivalent to several billion light-years. The mean redshift of a short GRB is about $z\approx0.5$. This cosmological distance propagation can accumulate small LV effects to the measurable level, and it is beneficial to our exploration of possible LV effects. GRBs are widely considered to consist of two phases: prompt phase and afterglow phase. The prompt phase usually lasts for tens of seconds and releases a large amount of total energy (${10}^{51} \mathrm{ergs}$\footnote{$\mathrm{erg}$ is a unit of energy and mechanical work, $1 \mathrm{erg}=10^{-7} \mathrm{J}=6.24\times10^{20} \,\mathrm{eV}$.}), and the prompt phase is followed by an afterglow, which lasts anywhere from a few weeks to a year.

Ref. \cite{C50-wei-2021-testing} divided the process of using GRB data for LV studies into three stages: \linebreak(i) Pre-Fermi studies stage: Ellis et al. are pioneers in this area of work \cite{C65-ellis-2003-quantum}. (ii) Fermi Large Area Telescope (Fermi/LAT) stage: due to the unprecedented sensitivity of Fermi/LAT for detecting the high-energy GRB emissions (up to tens of GeV), more stringent constraints on LV have been obtained using Fermi observations. (iii) The Major Atmospheric Gamma Imaging Cherenkov telescope (MAGIC) stage: recently, the MAGIC telescope has detected GRB emission at sub-TeV energy ranges, and MAGIC collaboration searched for the energy dependence of the arrival times of the sub-TeV scale photons and presented competitive limitations on the quadratic leading-order LV-induced vacuum dispersion \cite{C13-acciari-2020-bounds}. Besides these three stages, the commissioning of LHAASO will bring GRB detection into a new stage \cite{C66-lhaaso-2021-exploring}.

In the early research stages of using GRB data to limit LV parameters, people used one photon event from single GRB source to get many very strict limitations, but this method ignores the effect of intrinsic time delay from the source. Ref. \cite{C67-ellis-2006-robust} pointed out that any limitation from a single source can only be regarded as indicative limitation, since the limitation from a single source can be influenced by the unknown systematic uncertainties associated with the unknown intrinsic spectral properties of any given transient source. Ellis et al. used multiple GRB photon events to analyze LV effects \cite{C65-ellis-2003-quantum,C67-ellis-2006-robust,C68-ellis-2008-corrigendum}, and they used wavelet techniques to identify genuine features of high-energy photon light curve and that of low-energy photon, then they established a statistically robust lower limitation: $E_\mathrm{LV}>1.4\times10^{25}\,\mathrm{eV}$ \cite{C67-ellis-2006-robust,C68-ellis-2008-corrigendum}.

To account for the intrinsic time delay, Ellis et al. assumed a constant $b_\mathrm{sf}$ for the intrinsic time delay due to source effects \cite{C67-ellis-2006-robust}. Therefore, the observed delay of arrival time has two contributions $\Delta t_\mathrm{obs}=\Delta t_\mathrm{LV}+b_\mathrm{sf}(1+z)$, that reflects possible LV effects and source intrinsic effects \cite{C67-ellis-2006-robust}. Considering only the linear ($n=1$) LV modification in the subluminal case ($s_\mathrm{\pm}=+1$), and rescaling $\Delta t_\mathrm{obs}$ by the factor $(1+z)$, we can get a simple linear fit function:
\begin{linenomath}
\begin{equation}\label{GRB-linear fit}
\frac{\Delta t_\mathrm{obs}}{1+z}=a_\mathrm{LV}K+b_\mathrm{sf},
\end{equation}
\end{linenomath}
where
\begin{linenomath}
\begin{equation}\label{GRB-K}
K=\frac{1}{1+z}\int_0^z\frac{(1+z')dz'}{h(z')}
\end{equation}
\end{linenomath}
is a function of redshift, and this function depends on the cosmological model. The slope $a_\mathrm{LV}=\Delta E/(H_\mathrm{0}E_\mathrm{LV})$ is related to the LV scale, and the intercept $b_\mathrm{sf}$ denotes the possible unknown intrinsic time delay. In the standard flat $\Lambda\mathrm{CDM}$ model, the dimensionless expansion rate $h(z)$ is expressed as $h(z)=\sqrt{\Omega_\mathrm{m}(1+z)^3+\Omega_{\Lambda}}$. Refs. \cite{C69-biesiada-2009-lorentz,C70-pan-2015-constraints} extended this analysis to different cosmological models and showed that the result is insensitive to the adopted background cosmology. Subsequently, some cosmology-independent approaches were applied to probe the possible LV effects \cite{C71-zou-2018-model,C72-pan-2020-model}. The research of time delay effect is not sensitive to the cosmological parameters.

Actually, the traditional method, which uses only one GRB photon event without considering the intrinsic effects, is equivalent to only using the slope between a specific photon event and the point of origin. However, since the existence of internal time delay is not considered, using different GRB photon events will give very different results \cite{C51-7-shao-2010-lorentz,C52-zhang-2015-lorentz}. In the research method of Ellis et al. \cite{C65-ellis-2003-quantum,C67-ellis-2006-robust,C68-ellis-2008-corrigendum}, the intrinsic time delay in the linear fitting function is a constant $b_\mathrm{sf}$, so this method is equivalent to assume that all GRB high-energy photons have the same intrinsic time delay. However, since GRBs vary in duration, it is a question whether all high-energy photons radiated from different GRBs (or the same burst) have the same inherent time delay.

As an improvement, Zhang and Ma fitted the high-energy photon events from GRBs on several straight lines with the same slope as $1/E_\mathrm{LV,n}^n$ but with different intercepts $t_\mathrm{in}$ (different intrinsic emission times) \cite{C52-zhang-2015-lorentz}. This approach means that these photon events can be classified into different groups\footnote{Earlier Shao and Ma discussed the difference between long GRBs and short GRBs \cite{C51-7-shao-2010-lorentz}.}. Photon events between different groups have different internal time delay (that is, the intercepts of different lines are different), and this property means that photon events between different groups may have different emission mechanisms. However, photon events within the same group have the same internal time delay (that is, the intercept on the same line are same), and this property means that photon events between same groups may have similar emission mechanisms. All photon events have the same slope, that is, all photons follow the same LV laws. Note that there are three lines with different intercepts in Ref. \cite{C52-zhang-2015-lorentz}, but the photon events on the different lines are not evenly distributed, and the middle line (which is called the main line) has significantly more photon events than the other two lines. With the increase of data and the improvement of research methods, the main line shows greater research advantage \cite{C53-xu-2016-light-2}, and this fact implies that the emission times of different energy photon events at the source may have similar laws relative to low-energy photons.

The method of fitting a straight line by multiple GRB photon events can also be used to account for linear and quadratic modification researches. At early stages, the GRB data were not enough, so it is difficult to know which LV modification is dominant \cite{C51-7-shao-2010-lorentz,C37-shao-2010-lorentz,C52-zhang-2015-lorentz}. With the increase of data, Ref. \cite{C52-zhang-2015-lorentz} expressed the tendency, that the linear modification is dominant. Then by analyzing the data of high-energy photon events from seven GRBs with known redshift, Ref. \cite{C54-xu-2016-light} explicitly expressed that the linear modification is more dominant than the quadratic modification, and this characteristic becomes more and more obvious with the increase of data \cite{C53-xu-2016-light-2}.

Earlier the trigger time is often used as the characteristic time of low-energy photons~\cite{C51-7-shao-2010-lorentz,C52-zhang-2015-lorentz}, but the trigger time of low-energy photons is not only affected by the quantity of the detected low-energy photons but also limited by detector accuracy, so the trigger time may not be an objective standard to the characteristic time of low-energy photons \cite{C54-xu-2016-light}. Since peak time of light curve of low-energy photons can serve as a significant benchmark, and peak time is related to the intrinsic mechanisms of the GRBs objectively and naturally, so choosing peak time as the signal of the low-energy photons is more reasonable \cite{C54-xu-2016-light}. By comparing the light curves, the energy curves and the average energy curves of 8 GRBs with fixed redshift, Ref. \cite{C55-liu-2018-light} offered a new criterion, that use both the light curve and the average energy curves to determine the characteristic time for low-energy photons.

The approach with inherent time delay of GRB analysis could also lead to new insights into GRB emission mechanisms. We note that the intercept of the main line is negative (i.e., $b_\mathrm{sf}<0$), this property means that most of the high-energy photons are emitted earlier than the low-energy photons \cite{C51-7-shao-2010-lorentz,C37-shao-2010-lorentz,C52-zhang-2015-lorentz,C54-xu-2016-light,C53-xu-2016-light-2}, and this conclusion contradicts current general opinion. From the perspective of observation time, it is traditionally believed that high-energy photons start later than low-energy photons. One reason of the traditional opinion is that we usually assume that high-energy photons are generated by protons while low-energy photons are produced by electrons in source. Since electrons have lighter masses, they are accelerated earlier, and hence low-energy photons come out earlier \cite{C51-7-shao-2010-lorentz}. However, the emission mechanisms of GRBs are not complete, so this conclusion, high-energy photons start later than low-energy photons, may be changed. If we consider the delay of high-energy photons, the high-energy photons may be emitted earlier than the low-energy photons \cite{C51-7-shao-2010-lorentz,C37-shao-2010-lorentz,C52-zhang-2015-lorentz,C54-xu-2016-light,C53-xu-2016-light-2}. This view is also supported by early GRB models, such as GRB fireball model (for reviews see Refs. \cite{C126-piran-1994-fireballs,C127-meszaros-1995-gamma}), which asserts that GRBs are from fireballs. As a relativistic expanding plasma of electrons, photons and protons, the fireball becomes cooler and cooler during the expansion. At ultrahigh temperatures, photons cannot freely escape from the intense plasma due to frequent collisions and reactions, but with reduction of temperature, photons can run out and propagate outward. Because high-energy photons have higher energy, they can escape more easily, while low-energy photons can only escape after the fireball is further cooled, and this reason causes that the emission time of high-energy photons is earlier than that of low-energy photons. To further test this point, \mbox{Ref. \cite{C56-10-zhu-2021-pre}} found four photon events, which are high-energy photons with relatively low energies and have relatively small time delay effects, and these four photons arrive Earth earlier than low-energy photons, so these four events are direct signals of that high-energy photons can arrive Earth earlier. Through the method of machine learning, Ref. \cite{C57-12-chen-2021-novel} clearly put forward the point of view that there is an advanced stage, which is a stage of early emission of high-energy photons before the traditionally considered prompt and afterglow stages.

Through the above analysis, we can see that the approach with inherent time delay can provide stable support for LV researches, and through continuous improvement and development, so there is still a great possibility to provide meaningful research in the future. By analyzing 25 bright GRBs, Ref. \cite{C58-xu-2018-regularity} provided a general analysis on the reliability of LV studies using GRB data, and the results show that with the increase in the energies of photons, the regularity of LV truly emerges and gradually becomes significant. For photons with energies higher than $4\times10^{10}\,\mathrm{eV}$, the regularity exists at a significance of $3-5\sigma$ with $E_\mathrm{LV} = 3.6 \times 10^{26}\,\mathrm{eV}$ determined by the GRB data. Ref. \cite{C86-13-li-2021-light} pointed out that this modification of the speed of light obtained from GRB data can serve as key supporting evidence for string theory. Finally, we need to emphasize that the regularities revealed in the GRB data could be a consequence of the matter effects in space, and this possibility needs more researches.

Besides pure photon sector, linking high-energy neutrinos and GRBs could also be used for LV researches. Since neutrinos can escape from dense astrophysical environments and are not absorbed by background photons in the Universe, neutrinos can travel greater distances in the Universe with higher energies. Current GRB models predict the bursts of neutrinos with energies higher than $10^{14}\,\mathrm{eV}$ \cite{C75-waxman-1997-high,C96-vietri-1998-energy}, and comparisons of the neutrinos with the low-energy GRB photons could be used to test LV \cite{C97-amelino-2003-proposal,C98-jacob-2007-neutrinos}. Linking high-energy neutrinos and GRB photons requires consideration of the arrival times and orientations of high-energy neutrinos, and IceCube observatory\footnote{IceCube \cite{C113-achterberg-2006-first} is the largest neutrino detector in the world, IceCube is located in the ice near the South Pole, Antarctica, at a depth of $1.5\sim2.5\,\mathrm{km}$, and IceCube can detect neutrino sources up and down. IceCube neutrino observatory comprises three distinct components: a large buried array for ultrahigh energy neutrino detection, a surface air shower array, and a buried component called DeepCore. The first sensor was deployed during the austral summer of 2004–2005 and IceCube have been producing data since February 2005 \cite{C113-achterberg-2006-first}. All sensors were completed in 2010 and IceCube has continued to generate data \cite{C114-abbasi-2012-design}. The IceCube is designed to detect high-energy neutrinos emitted by extremely intense cosmic sources, such as black holes, exploding stars and neutron stars. When neutrinos collide with water molecules in the ice, they release high-energy subatomic particles, which move very quickly and emit short-lived light called Cherenkov radiation, and the radiation can be captured by IceCube, then IceCube can reconstruct the neutrino's route and identify its source.} can provide data support for this. Ref. \cite{C99-amelino-2017-vacuo} showed that the dispersion relation modifications for high-energy photons and high-energy neutrinos are roughly compatible. We should note that this property is not taken for granted, as we have mentioned earlier that the LV effect of different particles can be different. Ref. \cite{C103-18-huang-2018-lorentz} pointed out that such consistency between photons and neutrinos means positive support for the energy dependent speed variation of ultra-relativistic particles. Different from the speed modification characteristics of photons, neutrinos can be either superluminal or subluminal \cite{C103-18-huang-2018-lorentz}, and this property implies that the LV effect is a more plausible explanation than the material effect. One possible interpretation of this property is that neutrinos and anti-neutrinos may have superluminal and subluminal modification respectively, this interpretation means that there is an asymmetry between matter and antimatter, and this property can be explained by the CPT-odd feature of the linear LV \cite{C103-18-huang-2018-lorentz}. However, since IceCube cannot distinguish neutrinos and anti-neutrinos, this conjecture needs more research in support. Similar to the study of photons, the neutrino research also involves the effects of inherent time delay, and some research found that neutrinos are emitted earlier than GRB photons \cite{C103-18-huang-2018-lorentz,C104-20-huang-2019-consistent,C105-21-zhu-2021-pre}. Ref. \cite{C105-21-zhu-2021-pre} pointed out that there is a pre-burst stage, which emits neutrinos together with high-energy photons, and this pre-burst stage is within a time interval of about 140$\sim$820\,$\mathrm{s}$ before the low-energy photon burst of a GRB. Currently, the data of neutrinos are still relatively lacking, and there are prospects for development of LV research using neutrinos.

\subsubsection{Active Galactic Nuclei}\label{Active Galactic Nuclei}

AGN (for a review see Ref. \cite{C135-peterson-1997-introduction}) is a compact region at the center of a galaxy. AGNs contain quasars, seyfert galaxies, narrow line X-ray galaxies, low ionization nucleus, etc. AGN has a much higher than normal luminosity over some or all the electromagnetic spectrum, and its fast flux variations reach hours to years. Compared to GRBs, AGNs are closer to Earth, and their time structures are not so variable as those of GRBs. The gamma-rays of AGNs are higher in energy, and their very-high-energy ($E \ge10^{11}\,\mathrm{eV}$) emissions are in the TeV ranges, which are significantly higher than the highest energy observed in GRBs. TeV flares of AGNs have also been viewed as another kind of very-high-energy astronomical laboratories to probe possible evidence of LV.

There are some constraints on LV using observations of bright AGN flares, such as:
\begin{itemize}
\item	Whipple analysis of the flare of Mrk 421 ($z=0.031$) \cite{C59-biller-1999-limits}. Near the peak of the 15 May 1996 TeV flare from Markarian 421, there is no time delay larger than 280s between energy bands $< 1\,\mathrm{TeV}$ and $> 2\,\mathrm{TeV}$, and this result lead to a lower bound on the LV scale of $4\times10^{25}\,\mathrm{eV}$.
\item	MAGIC analysis of the flare of Mrk 501 ($z=0.034$) \cite{C60-albert-2008-probing}. During the very-high-energy flare of Mrk 501 between May and July 2005, there is a time delay about 4 mins for photons between energy bands 1.2--10$\,\mathrm{TeV}$ and 0.25--0.6$\,\mathrm{TeV}$. This finding establishes lower limitations $E_\mathrm{LV,1}>0.21\times10^{27}\,\mathrm{eV}$, $E_\mathrm{LV,2}>0.26\times10^{20}\,\mathrm{eV}$.
\item	H.E.S.S. analysis of the flare of PKS 2155-304 ($z=0.116$) \cite{C61-aharonian-2008-limits,C77-falomo-1993-environment}. During the very-high-energy flare of the PKS 2155-304 on 28 July in 2006, there is no significant time delay. This finding sets lower limitations $E_\mathrm{LV,1}>7.2\times10^{26}\,\mathrm{eV}$ (MCCF method), $E_\mathrm{LV,1}>5.2\times10^{26}\,\mathrm{eV}$ (wavelet analysis), $E_\mathrm{LV,2}>1.4\times10^{18}\,\mathrm{eV}$ (MCCF method).
\end{itemize}

\textls[-5]{In 2010, Shao and Ma briefly reviewed the analyses of the three AGNs. They recalculated the potential LV scales and discussed the hints and limitations from AGNs~\cite{C51-7-shao-2010-lorentz}. They found: (i) The analysis of Mrk 421 can sets a lower boundary to LV scale $E_\mathrm{LV,1}>4.9\times10^{25}\,\mathrm{eV}$, $E_\mathrm{LV,2}>1.5\times10^{19}\, \mathrm{eV}$; (ii) In the analysis of Mrk 501, mean difference of the two bands is reported $\approx2 \,\mathrm{TeV}$, this difference can sets $E_\mathrm{LV,1}\sim1.2\times10^{26} \,\mathrm{eV}$, and this value is very close to the results from the global fits of GRBs; (iii) The analysis of PKS 2155-304 includes a not sufficiently significant time delay $\sim$20$\, \mathrm{s}$ between lightcurves of two different energy bands, whose mean difference is $1.0\, \mathrm{TeV}$ and mean quadratic difference is $2.0\, \mathrm{TeV}^2$, and these differences set the potential LV scale to be $E_\mathrm{LV,1}$$\sim$$2.6\times10^{27} \,\mathrm{eV}$, $E_\mathrm{LV,2}$$\sim$$9.1\times10^{19} \,\mathrm{eV}$, which are both about one magnitude larger than the robust values from GRBs.}

Currently, AGN data are inadequate to carry out robust analyses \cite{C18-ellis-2009-probing}. The sources of AGNs are very different and have different unbeknown intrinsic mechanisms. Moreover, TeV flares from AGNs seem relatively rare and unpredictable, and these flares are produced only occasionally by AGNs with restricted redshift. Conversely, AGNs can provide a complementary probe to the light-speed variation from LV effects due to the different redshift and energy ranges from GRBs. GRBs can be detected at very large distances (their redshift are up to $z$$\sim$8), but with very limited high-energy photons ($E>$ tens of $\mathrm{GeV}$). On the contrary, AGN flares can be detected at very-high-energy ranges, but with limited distances ($z\le0.5$).

If we assume that the time delay effect in both GRBs and AGNs are same, we can check the analyses of the three AGNs in a different way. By assuming that some high-energy photons and low-energy photons from AGNs are emitted at the same time (like the ones corresponding to the light curve peaks), Li and Ma predicted the AGN time delay effects by the determined LV scale from GRBs $E_\mathrm{LV}^\mathrm{GRB}$, and they compared this time delay effects with the observed time delay effects to verify the light speed variation \cite{C62-11-li-2020-light}. This study showed that: (i) There is no time delay in Markarian 421 because that the 280 s binning is too large; (ii) The $4\pm1$ minutes time delay found by Markarian 501 can support the determined LV scale from GRBs $E_\mathrm{LV}^\mathrm{GRB}$; (iii) For PKS 2155-304, whose light curve has many peaks, it is hard to say how they correspond to each other, but there are also some peaks (the last peak) support the determined LV scale from GRBs. This study showed that several phenomena related to the light curves of the three AGNs can serve as the signal to support the results determined from GRBs.

\subsubsection{Pulsars}\label{Pulsars}

\textls[-15]{The first LV limitation of using gamma-ray radiation from the Galactic pulsar (for reviews see Refs. \cite{C139-lyne-2012-pulsar}) is from the High-Energy Gamma-ray Experiment Telescope (EGRET) on the Compton Gamma-ray Observatory (CGRO) at higher than $2 \,\mathrm{GeV}$ energy range \cite{C78-kaaret-1999-pulsar}. Recently, the MAGIC Collaboration demonstrated the strictest limitation derived from pulsars by studying the observed radiation from Crab pulsars up to TeV energy range, and the results are $E_\mathrm{LV,1}>5.5\times10^{26}\, \mathrm{eV}$ (linear subluminal case), $E_\mathrm{LV,1}>4.5\times10^{26} \,\mathrm{eV}$ (linear superluminal case), $E_\mathrm{LV,2}>5.9\times10^{19}\, \mathrm{eV}$ (quadratic subluminal case), $E_\mathrm{LV,2}>5.3\times10^{19}\, \mathrm{eV}$ (quadratic superluminal case) \cite{C79-ahnen-2017-constraining}.}

Gamma-ray pulsars are the only stable candidate astrophysical sources for such time delay studies. Pulsars can be observed for longer periods of time, and this property is of benefit to the sensitivity to LV researches. Pulsars have precise periodic flux variations, and pulsars are many orders of magnitude closer to Earth than GRBs and AGNs. The timing of pulsars has been carefully studied across the electromagnetic spectrum, so it is easier to distinguish energy-dependent time delay caused by LV from inherent mechanisms. However, pulsars are more near to Earth than GRBs and AGNs, and this property makes pulsars difficultly to accumulate LV effects through the long distance propagations. So the current LV studies from pulsars are limited.

\subsection{Vacuum Birefringence}\label{Vacuum Birefringence}

If the LV effects cause that photons with different helicity have different dispersion relations, there will be an underlying astrophysical LV phenomenon\hl{---}vacuum birefringence. In this potential LV phenomenon, photons with right-handed and left-handed polarization have different velocities, and the polarization vector of the linear polarization plane will rotate. The rotation of the polarization vector will accumulate with the increase of cosmological propagation distances, so this LV effect can be probed by astrophysical polarization measurements.

By the model-independent method, the velocity of photons with right-handed and left-handed polarization can be described by the main Taylor series expansion \cite{C129-mitrofanov-2003-constraint}:
\begin{linenomath}
\begin{equation}\label{vacuum birefringence-leading order}
v_\mathrm{\pm}=c[1\pm \eta (\frac{\hbar \omega}{E_\mathrm{Pl}})^n],
\end{equation}
\end{linenomath}
where $n$ is the $n$th-order energy dependence, and $n=1$ is linear term and $n=2$ is quadratic term. $\pm$ represents the different polarization states of photons, and $\eta$ is a dimensionless constant characterizing the degree of LV effects. Here, we consider linearly polarized light from an astronomical source, and this kind light is the superposition of two monochromatic waves with opposite circular polarization. With the LV effect (i.e. $\eta \ne 0$), the right-handed and left-handed photons emitted simultaneously from the same source will have different velocities, and this property can result a rotation of the polarization vector in the plane of linear polarization. The rotation angle $d\phi$, with the differential propagation distance $dL_\mathrm{(z)}$, can be expressed as \cite{C129-mitrofanov-2003-constraint,C130-jacobson-2004-new,C131-gleiser-2001-astrophysical,C92-gambini-1999-nonstandard}:
\begin{linenomath}
\begin{equation}\label{vacuum birefringence-differential rotation angle}
d\phi=\eta(\frac{\omega L_\mathrm{Pl}}{c})^{n+1}\frac{dL_\mathrm{(z)}}{L_\mathrm{Pl}},
\end{equation}
\end{linenomath}
where $L_\mathrm{Pl}$ is the Planck length and $\omega$ is the frequency of the observed photon. If we consider the expansion of the Universe, we can get the LV rotation angle $\Delta\phi_\mathrm{LV}$ during propagation from the source at redshift $z$ to the observer \cite{C133-zhou-2021-constraints}:
\begin{linenomath}
\begin{equation}\label{vacuum birefingence-LV rotation angle}
\Delta\phi_\mathrm{LV}=\frac{\eta}{H_\mathrm{0}}(\frac{L_\mathrm{Pl}}{c})^n(2\pi \omega)^{n+1}F(z,n),
\end{equation}
\end{linenomath}
where $H_\mathrm{0}$ is the Hubble constant, and the function $F(z,n)$ depends on the cosmological model. For the flat $\Lambda\mathrm{CDM}$ model \cite{C133-zhou-2021-constraints}:
\begin{linenomath}
\begin{equation}\label{vacuum briefingence-F}
F(z,n)=\int_0^z\frac{(1+z')^ndz'}{\sqrt{\Omega_\mathrm{m}(1+z')^3+\Omega_{\Lambda}}},
\end{equation}
\end{linenomath}
where $\Omega_\mathrm{m}$ and $\Omega_{\Lambda}$ are the matter energy density and vacuum energy density. Only considering linear term ($n=1$), we have \cite{C12-toma-2012-steict,C10-laurent-2011-constraints}:
\begin{linenomath}
\begin{equation}\label{vacuum briefingence-LV rotation angle (linear order)}
\Delta\phi_\mathrm{LV}(E)=\eta\frac{E^2}{\hbar E_\mathrm{Pl}H_\mathrm{0}}\int_0^z\frac{(1+z')dz'}{\sqrt{\Omega_\mathrm{m}(1+z')^3+\Omega_{\Lambda}}}.
\end{equation}
\end{linenomath}

If we consider both the intrinsic polarization angle $\phi_\mathrm{in}$ caused by the source effects and the rotation angle $\Delta\phi_\mathrm{LV}$ caused by the LV effects, the observed polarization angle should contain two items \cite{C133-zhou-2021-constraints}:
\begin{linenomath}
\begin{equation}\label{vacuum briefingence-rotation angle (obs)}
\phi_\mathrm{obs}=\phi_\mathrm{in}+\Delta\phi_\mathrm{LV}(E)=\phi_\mathrm{in}+BE^2,
\end{equation}
\end{linenomath}
where $B=(\eta F(z))/(\hbar E_\mathrm{Pl}H_\mathrm{0})$.

Since the emission mechanisms of astronomical sources are still poorly understood, it is difficult to distinguish the intrinsic polarization angle from the LV-induced rotation angle. If the emission mechanisms are available, vacuum birefringence can be examined by measuring the observed polarization angle with the intrinsic polarization angle removed. Without these knowledge, vacuum birefringence can still be limited, because the rotation angle caused by LV effects can offset partially, but not all, polarization angle caused by emission mechanisms. Therefore, the astrophysical polarization measurements can put an upper limitation on this possible LV phenomenon.

Research in astrophysics has put strict limitations on vacuum birefringence. \mbox{Ref. \cite{C50-wei-2021-testing}} summarized the limitations we have obtained so far on vacuum birefringence, the strictest limitation was from Götz et al.: $\eta<1.0\times10^{-16}$, which is obtained from the data of farthest polarized burst (whose redshift is up to $z=2.739$) GRB 140206A \cite{C80-gotz-2014-GRB}. Besides GRB, gamma-rays from other celestial bodies can also provide constraints on the vacuum birefringence effect, such as Cygnus \cite{C148-Shao-2011-lorentz}. We need to note that the LV parameters can be process-dependent, and this property means that strict limitations on vacuum birefringence are not equivalent to strict limitations on all LV phenomena. In other words, even if we exclude the existence of vacuum birefringence, we cannot exclude the existence of other LV phenomena, and we cannot exclude the existence of LV effects.

The strict limitations of vacuum birefringence pose serious challenges to some theoretical models, especially, vacuum birefringence can serve to distinguish parity-violating theories from those of even parity \cite{C148-Shao-2011-lorentz}. These strict limitations on vacuum birefringence have challenged some theoretical models that allow the existence of vacuum birefringence, such as SME. However, other theories that do not have vacuum birefringence can avoid these limitations naturally, such as D-foam model in string theory \cite{C86-13-li-2021-light}.

Compared with the time delay measurement, the polarization measurement is more restricted to the LV effects. From Equations (\ref{time delay-LV delay}) and (\ref{vacuum briefingence-LV rotation angle (linear order)}), we can know that the energy dependence proportionality coefficient of time delay is $E$ when that of polarization is $E^2$. This property means that the polarization measurement is more sensitive to the photon energy $E$ than the time delay measurement. However, many theories do not have the LV predicted signals of vacuum birefringence, so the limitations from time delay measurements are essential in the extensive search for LV effects.

\subsection{Threshold Anomaly of Photon Annihilation Reaction}\label{Threshold anomaly of photon annihilation reaction}

According to the special relativity, the photon annihilation reaction $\gamma + \gamma_\mathrm{b} \to e^\mathrm{+}+e^\mathrm{-}$ can prevent cosmic photons, with energy higher than the threshold, from traveling gigantic distances in the Universe. This property can result an absorption modification of the spectrum, so we also call this annihilation reaction the background absorption of high-energy photons. $\gamma_\mathrm{b}$ is the background low-energy photons, which can be provided by the cosmic microwave background (CMB)\footnote{After the Big Bang, photons decoupled from matter and evolved as non-interacting particles with the expansion of the Universe, then these photons served as the CMB.} or by the extra-galactic background light (EBL)\footnote{In the process of evolution of the Universe, various luminous bodies emitted photons (mainly in the radio, infrared and other bands), which remained and pervaded the entire cosmic space, at last these photons became the EBL.}.

We can assume that the low-energy photon $\gamma_\mathrm{b}$ is a fixed photon, its 4-momentum is $p_\mathrm{2}$, and its energy $\varepsilon_\mathrm{b}$ and momentum $|\vec{p_\mathrm{2}}|$ are much smaller than that of high-energy photon $\gamma$ with 4-momentum $p_\mathrm{1}$. When the angle between the two incident photons is $\pi$, the angle between the outgoing electron-positron pair is $0$, and these two outgoing particles severally carry half the incident energy and momentum, we can take the threshold of this reaction. According to the special relativity, high-energy photon threshold of the annihilation reaction $\gamma + \gamma_\mathrm{b} \to e^\mathrm{+}+e^\mathrm{-}$ is:
\begin{linenomath}
\begin{equation}\label{photon reapp-reaction threshold (SR)}
E\ge E_\mathrm{th}=\frac{m^2_\mathrm{e}}{\varepsilon_\mathrm{b}},
\end{equation}
\end{linenomath}
and there is no upper threshold for this reaction. Once the energy of the high-energy photon exceeds the lower threshold, in the gigantic distance propagation of the Universe, there will be certain configurations that allow this reaction to occur. So this reaction will cause the high-energy photon to be absorbed by the background photons and unable to reach the Earth. That is to say, if the photons come from far enough away, we cannot observe photons with energies above the threshold on Earth. The higher the energy of the high-energy photon, the lower the required response threshold for background photons. In the corresponding electromagnetic band in the cosmic space, there are many low-energy photons, so the higher-energy photon is easier to be annihilated.

If the LV effects modify the dispersion relation of photons, the annihilation reaction between high-energy photons and low-energy photons will produce interesting physical phenomena. By model-independent method, the modified dispersion relation for photons can be expressed as:
\begin{linenomath}
\begin{equation}\label{photon reapp-dispersion relation}
\omega(k)^2=k^2-\xi k^n,
\end{equation}
\end{linenomath}
where $\xi$ is the LV parameter, which can be positive (subluminal), negative (superluminal) or zero. If the LV parameter $\xi$ takes a value in a specific interval, this annihilation reaction will show special phenomena. For example, the high-energy photons that should have been annihilated are not annihilated, and we can observe high-energy photons whose energy is above the threshold of the special relativity.

We assume that the 4-momentum of the high-energy photon $\gamma$ is $p_\mathrm{1}=(\omega(k),0,0,k)$ with LV effects, and that of the low-energy photon $\gamma_\mathrm{b}$ is still $p_\mathrm{2}=(\varepsilon_\mathrm{b},0,0,-\varepsilon_\mathrm{b})$\footnote{It is not necessary to modify the 4-momentum of the low-energy photon $\gamma_\mathrm{b}$, because $\varepsilon_\mathrm{b}$ is so small that the LV effects are negligible.}. According to the threshold configuration of this reaction\footnote{The threshold configuration is that: the angle between the two incident photons is $\pi$, the angle between the outgoing electron-positron pair is $0$, and these two outgoing particles severally carry half the incident energy and momentum.}, the energy-momentum conservation and the on-shell condition of the outgoing electron-positron pair, we get \cite{C38-li-2021-threshold}:
\begin{linenomath}
\begin{equation}\label{photon reapp-reaction condition}
m^2_\mathrm{e}=\frac{-\xi k^n+4k\varepsilon_\mathrm{b}}{4}+\mathcal{O}(\xi^2)+\mathcal{O}(\xi\varepsilon_\mathrm{b}),
\end{equation}
\end{linenomath}
equivalently:
\begin{linenomath}
\begin{equation}\label{photon reapp-xi(f)}
\xi=f(k)=\frac{4\varepsilon_\mathrm{b}}{k^{n-1}}-\frac{4m^2_\mathrm{e}}{k^n},\quad k>0.
\end{equation}
\end{linenomath}

A function $f(k)$ is defined in the above formula \cite{C38-li-2021-threshold}. Then we consider a simplified case ($n=3$), which corresponds to the linear modification. The function $f(k)$ has only one zero: $k_\mathrm{0}=(m^2_\mathrm{e})/(\varepsilon_\mathrm{b})$, which is the lower threshold derived in the case of special relativity. $f(k)$ tends to zero at $k=+\infty$, tends to $-\infty$ at $k=0$. $f(k)$ has the maximum value $\mathrm{max}f(k)=f(k_\mathrm{c})=(16\varepsilon^3_\mathrm{b})/(27m^4_\mathrm{e})$ at the critical point $k_\mathrm{c}=(3m^2_\mathrm{e})/(2\varepsilon_\mathrm{b})$. Now studying the threshold behavior is equivalent to studying the solution of Equation (\ref{photon reapp-xi(f)}), and we can acquire the number and locations of solutions of Equation (\ref{photon reapp-xi(f)}) by studying the relationship between the value of $\xi$ and the function $f(k)$. Finally, according to the three kinds different values of the LV parameter $\xi$, the threshold of the annihilation reaction $\gamma + \gamma_\mathrm{b} \to e^\mathrm{+}+e^\mathrm{-}$ has three different cases \cite{C38-li-2021-threshold}:
\begin{itemize}
\item	Case I. $\xi>\mathrm{max}f(k)$ (optical transparency)

Equation (\ref{photon reapp-xi(f)}) has no solution. No solution means that there is not even a lower threshold, and all photons can freely traverse the background of low-energy photons. Case I is a subluminal effect.
\item	Case II. $0\le\xi\le \mathrm{max}f(k)$ (reappearance of ultra-high-energy photons)\\
Equation (\ref{photon reapp-xi(f)}) has two different solutions, and the smaller and larger solutions are denoted by $k_\mathrm{<}$ and $k_\mathrm{>}$ respectively. Obviously, $k_\mathrm{<}$ is the lower threshold, but there is an additional solution $k_\mathrm{>}$. With the help of the theorem in Ref. \cite{C17-maccione-2007-constraints}, $k_\mathrm{>}$ can be determined as an upper threshold. Klu{\'z}niak pointed out the possibility of the upper threshold for two photons annihilating to electron-positron pair \cite{C134-kluzniak-1999-transparency}. In case II, only photons with energy between $\omega(k_\mathrm{<})$ and $\omega(k_\mathrm{>})$ can be absorbed by photons with energy $\varepsilon_\mathrm{b}$. Thus, we get an interesting conclusion: background photon with energy $\varepsilon_\mathrm{b}$ is optically transparent as usual to photon with energy lower than $\omega(k_\mathrm{<})$, while the background photon is transparent again for photon with energy higher than $\omega(k_\mathrm{>})$. This conclusion means that the high-energy photon with energy higher than $\omega(k_\mathrm{>})$ can arrival on Earth. In particular, when $\xi$ tends to zero, $k_\mathrm{>}$ tends to $+\infty$, this property means we go back to the case of special relativity. That is to be expected, since any theory must go back to classical theory in the low-energy ranges. Case II is also a subluminal effect.
\item	Case III. $\xi<0$ (threshold reduction)
Equation (\ref{photon reapp-xi(f)}) has only one solution, which is a lower threshold and is smaller than $k_\mathrm{0}$. The threshold behavior here is almost the same as in special relativity, except that the lower threshold is more lower. Case III is a superluminal effect.
\end{itemize}

In case III, the lower threshold for the photon annihilation reaction is more lower, so we need other processes (such as photon decay) to set stricter limitations on $\xi$. For case I and case II, if we assume that the low-energy photon with 4-momentum $p_\mathrm{2}$ comes from the CMB, then we can get that the average energy of low-energy photons is $\varepsilon_\mathrm{b}\simeq6.35\times10^{-4} \,\mathrm{eV}$, and therefore we get $k_\mathrm{c}=3/2k_\mathrm{0}\simeq6.17\times 10^{14} \,\mathrm{eV}$ \cite{C38-li-2021-threshold}. According to the special relativity, if the energy of a photon exceeds $k_\mathrm{0}\simeq4.11\times 10^{14} \,\mathrm{eV}$, the CMB has a significant absorption effect on this photon. If we observe a disproportionate number of photons with energies over $k_\mathrm{0}\simeq4.11\times 10^{14} \,\mathrm{eV}$ and link them to distant extragalactic sources, we can attribute these observations as support to LV effects.

The ground-based cosmic-ray observatory, such as LHAASO, can be an ideal platform for such observations \cite{C38-li-2021-threshold}. Ref. \cite{C81-14-li-2021-ultrahigh} analyzed the high-energy photon datum of $1.42\, \mathrm{PeV}$~\cite{C88-cao-2021-ultrahigh}, newly discovered by LHAASO, from the perspective of photon annihilation reaction threshold anomaly, and then Refs. \cite{C81-14-li-2021-ultrahigh,C38-li-2021-threshold,C151-Li-2021-lhaaso} proposed to search for PeV scale photons from extragalactic sources as a strong sign of subluminal LV\footnote{The high-energy photon of $1.42\, \mathrm{PeV}$ currently observed by LHAASO is affected by the free path of photons, the source of photons, and the energy of background photons, so it cannot yet be directly considered to be the exact signal of LV-induced photon annihilation reaction threshold anomaly, but this $1.42 \,\mathrm{PeV}$ photon can be seen as a positive signal for LV effect.}.

\subsection{Decay of High-Energy Photons}\label{Decay of High-Energy Photons}

In the standard model, limited by the energy-momentum conservation, the photon decay reaction $\gamma\to e^\mathrm{+}+e^\mathrm{-}$ is prohibited, but in the theories including LV effects, photon decay may be a possible phenomenon \cite{C39-jacobson-2003-threshold}. By model-independent method, the energy-momentum dispersion relation of photons can be expressed as:
\begin{linenomath}
\begin{equation}\label{photon decay-dispersion relation (photon)}
\omega^2=k^2[1+\xi_\mathrm{n}(\frac{k}{E_\mathrm{Pl}})^n],
\end{equation}
\end{linenomath}
that of electrons is:
\begin{linenomath}
\begin{equation}\label{photon decay-dispersion relation (electron)}
E^2=m^2+p^2[1+\eta_\mathrm{n}(\frac{p}{E_\mathrm{Pl}})^n],
\end{equation}
\end{linenomath}
where $\xi_\mathrm{n}$, $\eta_\mathrm{n}$ are the $n$th-order LV parameters of photons and electrons, respectively. For a specific theory, we can choose appropriate parameters $\xi_\mathrm{n}$, $\eta_\mathrm{n}$ to be consistent with the theory.

Considering a high-energy photon with momentum $k$ that decays into an electron with momentum $xk$ ($x\in[0,1]$) and a positron with momentum $(1-x)k$, using \mbox{Equation (\ref{photon decay-dispersion relation (photon)})}, Equation (\ref{photon decay-dispersion relation (electron)}) and the energy-momentum conservation relation, and expanding to the first-order of the LV parameters and the leading-order of $(m/k)^2$, we can get \cite {C39-jacobson-2003-threshold}:
\begin{linenomath}
\begin{equation}\label{photon decay-reaction condition}
k[1+\frac{\xi_\mathrm{n}}{2}(\frac{k}{E_\mathrm{Pl}})^n]=xk[1+\frac{m^2}{2(xk)^2}+\frac{\eta_\mathrm{n}}{2}(\frac{xk}{E_\mathrm{Pl}})^n]+\{x\leftrightarrow 1-x\}.
\end{equation}
\end{linenomath}

After simple algebraic operations, the above formula becomes \cite{C39-jacobson-2003-threshold}:
\begin{linenomath}
\begin{equation}\label{photon decay-reaction condition (simple)}
\frac{m^2E^n_\mathrm{Pl}}{k^{n+2}}=x(1-x)[\xi_\mathrm{n}-\eta_\mathrm{n}((1-x)^{n+1}+x^{n+1})].
\end{equation}
\end{linenomath}

Equation (\ref{photon decay-reaction condition (simple)}) means that finding the threshold of the photon decay reaction is equivalent to finding the minimum value of $k$ on the left side of the equation, and correspondingly, maximizing the right side of the equation. In the follow discussion, we only consider linear modifications and quadratic modifications by situations:\\
\begin{itemize}
\item	Case I. $\xi_\mathrm{1} \ne 0, \xi_\mathrm{2}=\eta_\mathrm{1}=\eta_\mathrm{2}=0$\\
This parameter selection corresponds to only a photon linear LV modification, but no other LV modifications. Then Equation (\ref{photon decay-reaction condition (simple)}) becomes \cite{C37-shao-2010-lorentz}:
\begin{linenomath}
\begin{equation}\label{photon decay-reaction condition (simple,I)}
m^2E_\mathrm{Pl}/k^3=[x(1-x)]\xi_1.
\end{equation}
\end{linenomath}
\begin{enumerate}
\item When $\xi_\mathrm{1}\to0$, $k\to+\infty$, it is the case where photons cannot decay in the standard model;
\item	When $\xi_\mathrm{1}<0$, there is no photon decay \cite{C169-Jacobson-2001-TeV,C39-jacobson-2003-threshold};
\item	When $\xi_\mathrm{1}>0$, the threshold of photon decay is \cite{C39-jacobson-2003-threshold}:\\
\begin{linenomath}
\begin{equation}\label{photon decay-reaction threshold (I)}
k_\mathrm{th}=(\frac{4m^2E_\mathrm{Pl}}{\xi_\mathrm{1}})^{1/3}.
\end{equation}
\end{linenomath}
\end{enumerate}

In case I, photon decay means that the LV effects only result in a linear superluminal modification for photons. Assuming that the photon superluminal modification energy scale is $E_\mathrm{LV}^\mathrm{(sup)}=\xi_\mathrm{1}^{-1}$, then from the above formula we can obtain \cite{C81-14-li-2021-ultrahigh}:
\begin{linenomath}
\begin{equation}\label{photon decay-LV scale (I)}
E_\mathrm{LV}^\mathrm{(sup)}\ge9.57(\frac{E_\gamma}{\mathrm{PeV}})^3\times10^{32} \mathrm{eV},
\end{equation}
\end{linenomath}
where $E_\gamma$ is the energy of high-energy photon.\\

\item	Case II. $\xi_\mathrm{1}=\eta_\mathrm{1} \ne 0, \xi_\mathrm{2}=\eta_\mathrm{2}=0$\\
This parameter selection corresponds to a simple assumption: the LV modification parameters of photons and electrons are same, and both are linear modifications. Then Equation (\ref{photon decay-reaction condition (simple)}) becomes \cite{C37-shao-2010-lorentz}:
\begin{linenomath}
\begin{equation}\label{photon decay-reaction condition (simple,II)}
m^2E_\mathrm{Pl}/k^3=2[x^2(1-x)^2]\xi_\mathrm{1}.
\end{equation}
\end{linenomath}
\begin{enumerate}
\item	When $\xi_\mathrm{1}=\eta_\mathrm{1}<0$, there is no photon decay \cite{C37-shao-2010-lorentz};
\item	When $\xi_\mathrm{1}=\eta_\mathrm{1}>0$, the threshold of photon decay is \cite{C37-shao-2010-lorentz}:
\begin{linenomath}
\begin{equation}\label{photon decay-reaction threshold (II)}
k_\mathrm{th}=(\frac{8m^2E_\mathrm{Pl}}{\xi_\mathrm{1}})^{1/3}.
\end{equation}
\end{linenomath}
\end{enumerate}

In case I and case II, the threshold of photon decay occurs at $x=1/2$, that is, the momenta of the electron-positron pair generated by photon decay are equally distributed; however,  this is not always the case \cite{C169-Jacobson-2001-TeV,C39-jacobson-2003-threshold,C37-shao-2010-lorentz}.\\

\item Case III. $\xi_\mathrm{1}\ne0, \eta_\mathrm{1}\ne0, \xi_\mathrm{2}=\eta_\mathrm{2}=0$\\
This parameter selection means that neither photons nor electrons has quadratic modifications, but they have different linear modification terms. Then Equation (\ref{photon decay-reaction condition (simple)}) becomes \cite{C39-jacobson-2003-threshold}:
\begin{equation}\label{e-m relation of photon decay for linear modification, eta/=0, xi/=0}
\frac{m^2E_\mathrm{Pl}}{k^3}=x(1-x)[\xi_\mathrm{1}-((1-x)^2+x^2)\eta_\mathrm{1}].
\end{equation}

Introducing a new variable $z=(2x-1)^2$ can make the analyse simple, so that $x=(1+\sqrt{z})/2$, $(1-x)=(1-\sqrt{z})/2$ and $x(1-x)=(1-z)/4$. The relevant range of $z$ is $[0,1]$, where $z=0$ corresponds to the symmetric configuration $x=1/2$ and $z=1$ corresponds to $x=1$ \cite{C39-jacobson-2003-threshold}. In terms of $z$, Equation (\ref{e-m relation of photon decay for linear modification, eta/=0, xi/=0}) becomes \cite{C39-jacobson-2003-threshold}:
\begin{equation}\label{e-m relation of photon decay for linear modification, eta/=0, xi/=0, z}
\frac{m^2E_\mathrm{Pl}}{k^3}=\frac{1-z}{2}\xi_\mathrm{1}-\frac{1-z^2}{8}\eta_\mathrm{1}.
\end{equation}
\begin{enumerate}
\item	When $\xi_\mathrm{1}<\eta_\mathrm{1}$ and $2\xi_\mathrm{1}-\eta_\mathrm{1}<0$, there is no photon decay \cite{C169-Jacobson-2001-TeV,C39-jacobson-2003-threshold};
\item	When $\xi_\mathrm{1}>0$ and $2\xi_\mathrm{1}-\eta_\mathrm{1}>0$, the threshold of photon decay is \cite{C169-Jacobson-2001-TeV,C39-jacobson-2003-threshold}:
\begin{equation}\label{threshold of photon decay, xi>0, 2xi-eta>0}
k_\mathrm{th}=(\frac{8m^2E_\mathrm{Pl}}{2\xi_\mathrm{1}-\eta_\mathrm{1}})^{1/3}.
\end{equation}
This threshold is taken at $z=0$, that is $x=1/2$, the momenta of the out-going particles generated by photon decay are equally distributed;
\item	When $\eta_\mathrm{1}<\xi_\mathrm{1}<0$, the threshold of photon decay is \cite{C169-Jacobson-2001-TeV,C39-jacobson-2003-threshold}:
\begin{equation}\label{threshold of photon decay, eta<xi<0}
k_\mathrm{th}=(\frac{-8\eta_\mathrm{1}m^2E_\mathrm{Pl}}{(\xi_\mathrm{1}-\eta_\mathrm{1})^2})^{1/3}.
\end{equation}
This threshold is taken at $z=\xi_\mathrm{1}/\eta_\mathrm{1}$, that is $x=(1+\sqrt{\xi_\mathrm{1}/\eta_\mathrm{1}})/2$.\\
\end{enumerate}

\item	Case IV. $\xi_\mathrm{1}=\eta_\mathrm{1}=0, \eta_\mathrm{2}\ne0, \xi_\mathrm{2}\ne0$\\
This parameter selection means that neither photons nor electrons has linear modifications, but they have different quadratic modification terms. Then Equation (\ref{photon decay-reaction condition (simple)}) becomes \cite{C37-shao-2010-lorentz}:
\begin{linenomath}
\begin{equation}\label{photon decay-reaction condition (simple,III-1)}
\frac{m^2E^2_\mathrm{Pl}}{k^4}=x(1-x)[\xi_\mathrm{2}-\eta_\mathrm{2}((1-x)^3+x^3)].
\end{equation}
\end{linenomath}
If $\eta_\mathrm{2}=0$, the situation becomes:
\begin{enumerate}
\item	When $\xi_\mathrm{2}<0$, there is no photon decay \cite{C37-shao-2010-lorentz};
\item	when $\xi_\mathrm{2}>0$, the threshold of photon decay is \cite{C37-shao-2010-lorentz}:
\begin{linenomath}
\begin{equation}\label{photon decay-reaction threshold (III-1)}
k_\mathrm{th}=(\frac{4m^2E^2_\mathrm{Pl}}{\xi_\mathrm{2}})^{1/4}.
\end{equation}
\end{linenomath}
\end{enumerate}

For the case of $\eta_\mathrm{2}\ne0$, introducing $z=(2x-1)^2$ ($z\in[0,1]$) can make that the above fourth power Equation (\ref{photon decay-reaction condition (simple)}) about $x$ becomes the quadratic power equation about $z$~\cite{C39-jacobson-2003-threshold}:
\begin{linenomath}
\begin{equation}\label{photon decay-reaction condition (simple,III-2)}
\frac{m^2E^2_\mathrm{Pl}}{k^4}=\frac{3\eta_\mathrm{2}}{16}(z-\frac{2\xi_\mathrm{2}+\eta_\mathrm{2}}{3\eta_\mathrm{2}})^2-\frac{1}{12\eta_\mathrm{2}}(\xi_\mathrm{2}-\eta_\mathrm{2})^2.
\end{equation}
\end{linenomath}
After detailed calculation, we get:
\begin{enumerate}
\item When $\xi_\mathrm{2}<\eta_\mathrm{2}$, and $4\xi_\mathrm{2}<\eta_\mathrm{2}$, there is no photon decay \cite{C39-jacobson-2003-threshold};
\item	When $-2\xi_\mathrm{2}<\eta_\mathrm{2}<4\xi_\mathrm{2}$, the reaction threshold of photon decay is \cite{C39-jacobson-2003-threshold}:
\begin{linenomath}
\begin{equation}\label{photon decay-reaction threshold (III-2)}
k_\mathrm{th}=(\frac{16m^2E^2_\mathrm{Pl}}{4\xi_\mathrm{2}-\eta_\mathrm{2}})^{1/4}.
\end{equation}
\end{linenomath}
This threshold is taken at $z=0$, that is $x=1/2$. In this case, the momenta are equally distributed;
\item	When $\xi_\mathrm{2}>\eta_\mathrm{2}$ and $-2\xi_\mathrm{2}>\eta_\mathrm{2}$, the reaction threshold of photon decay is \cite{C39-jacobson-2003-threshold}:
\begin{linenomath}
\begin{equation}\label{photon decay-reaction threshold (III-3)}
k_\mathrm{th}=(\frac{-12\eta_\mathrm{2}m^2E^2_\mathrm{Pl}}{(\xi_\mathrm{2}-\eta_\mathrm{2})^2})^{1/4}.
\end{equation}
\end{linenomath}
This threshold is taken at $z=(2\xi_\mathrm{2}+\eta_\mathrm{2})/3\eta_\mathrm{2}$, that is \linebreak $x=[1\pm\sqrt{(2\xi_\mathrm{2}+ \eta_\mathrm{2})/3\eta_\mathrm{2}}]/2$ \cite{C39-jacobson-2003-threshold}. In this case, the momenta are not equally distributed.
\end{enumerate}

\end{itemize}

At present, the astronomical observations have obtained strict limitations on the photon decay \cite{C81-14-li-2021-ultrahigh,C66-lhaaso-2021-exploring}. Ref. \cite{C81-14-li-2021-ultrahigh} inserted the date $E_{\gamma\mathrm{(max)}}=1.42 \,\mathrm{PeV}$ (the highest-energy event from LHAASO J2032+4102 \cite{C88-cao-2021-ultrahigh}) into Equation (\ref{photon decay-LV scale (I)}) and got $E_\mathrm{LV}^\mathrm{(sup)}\ge2.74\times10^{33} \,\mathrm{eV}$. This result is likely to be the strongest constraint from the LHAASO data. LHAASO collaboration conducted a pseudo-experiment by MC simulations and adopted the CLs method~\cite{C66-lhaaso-2021-exploring}. They got a precise $95\% \mathrm{CL}$ lower limitation on the cut-off energy of the spectrum of two sources LHAASO J2032+4102 and J0534+2202. The first-order superluminal LV scale is constrained to be higher than $10^5M_\mathrm{Pl}$, and the second-order superluminal LV scale should exceed $10^{-3}M_\mathrm{Pl}$. These results are the strongest constraints on the superluminal LV parameters among experimental results with similar technique\footnote{Ref. \cite{C81-14-li-2021-ultrahigh} only used the highest-energy photon datum, while LHAASO collaboration used the data of the whole spectrum \cite{C66-lhaaso-2021-exploring}. The limitation using only the highest-energy photon is tighter, and the limitation using the entire spectrum is more robust and reliable.}. Note that only considering the case I means that the LV effects only result a linear superluminal modification on photons, and the strict limitation on the photon decay can also be regarded as a strict limitation on the superluminal modification. When we study photon decay, we usually only consider case I.

\section{Summary}\label{Summary}

Current astronomical observations provide rich data for LV research and provide various constraints on LV effects. For cosmic photons, the time delay effect obtained from research on the GRB data, suggests a linear subluminal modification for the speed of light, and this finding can be considered as an important supportive signal for LV of photons. Strict limitations of vacuum birefringence mean that the modification of the speed of light is independent of the helicity of photons. The currently observed ultra-high-energy photon signals may be examples of the photon annihilation reaction threshold anomaly, and these signals may be another positive supports for the subluminal modification of photon velocity. However, the strict limitations on photon decay phenomenon impose strict constraints on the superluminal modification of photon velocity.

Different theories have different LV predictions, and this property means that we can verify or constrain different theories by different limitations from different LV phenomena. The linear subluminal energy-dependent modification of photon velocity, which is obtained from photon time delay effect, is supportive for some theoretical models. The strict limitations of vacuum birefringence mean that some theories, which support the helicity dependence of the photon velocity, are severely challenged. The strict limitations of photon decay mean that some theories, which support the superluminal modification, are severely limited. The D-foam model in string theory does not contain the helicity dependence of the photon velocity, so this model can naturally avoid the strict limitations brought about by the vacuum birefringence; this model does not support the superluminal energy dependence modification of photon velocity, so this model can also avoid the strict limitations brought about by photon decay; and this model has obtained positive supports from time delay effect and threshold anomaly of photon annihilation reaction. As a summary, this string theory model for space-time foam is a viable theory with Lorentz violation of photons.

We should note that the LV effects can be particle-dependent and process-dependent. This property means that strict constraints obtained from a certain process of a certain particle do not refute the full possibility of LV with other processes or other particles. Specific processes from specific particles can provide supportive evidences for LV effects, so the existence of LV might take a long time to verify along with consistently ongoing progresses of LV research both experimentally and theoretically.

\vspace{6pt}

\authorcontributions {P.H. and B.-Q.M. equally contributed to all stages of this project. All authors have read and agreed to the published version of the manuscript.}

\funding{\textls[-15]{This work is supported by National Natural Science Foundation of China (Grant No. 12075003).}}

\conflictsofinterest{The authors declare no conflict of interest.}

\newpage
\abbreviations{Abbreviations}{
The following abbreviations are used in this manuscript:\\

\noindent
\begin{tabular}{@{}ll}
LV & Lorentz symmetry violation\\
DSR & Doubly special relativity\\
SME & Standard model extension\\
LHC & Large Hadron-Collider\\
LHAASO & Large High Altitude Air Shower Observatory\\
GRB & Gamma-ray burst\\
AGN & Active galactic nucleus\\
GZK & Greisen–Zatsepin–Kuzmin\\
Fermi/LAT & Fermi Large Area Telescope\\
MAGIC & Major Atmospheric Gamma Imaging Cherenkov telescope\\
EGRET & High-Energy Gamma-ray Experiment Telescope\\
CGRO & Compton Gamma-ray Observatory\\
CMB & Cosmic microwave background\\
EBL & Extra-galactic background light
\end{tabular}}


\begin{adjustwidth}{-\extralength}{0cm}

\printendnotes[custom]

\reftitle{References}




\end{adjustwidth}
\end{document}